\documentclass[11pt,a4paper]{article}
\usepackage[T1]{fontenc}
\usepackage[utf8]{inputenc}
\usepackage[a4paper,margin=1in]{geometry}
\usepackage{lmodern}
\usepackage{amsmath,amssymb,amsfonts}
\usepackage{mathtools}
\usepackage{bm}
\usepackage{graphicx}
\usepackage{xcolor}
\usepackage{booktabs}
\usepackage[font=small,labelfont=bf,labelsep=period,justification=justified]{caption}
\usepackage{authblk}
\usepackage{titlesec}
\usepackage{fancyhdr}
\usepackage[numbers,sort&compress]{natbib}
\usepackage[colorlinks=true,
            citecolor=black,
            linkcolor=black,
            urlcolor=black]{hyperref}
\usepackage{doi}
\graphicspath{{figures/}}
\renewcommand{\thesection}{\Roman{section}}
\renewcommand{\thesubsection}{\Alph{subsection}}
\titleformat{\section}
  {\normalfont\scshape\centering}
  {\thesection.}{0.6em}{}
\titleformat{\subsection}
  {\normalfont\bfseries}
  {\thesubsection.}{0.5em}{}
\titlespacing*{\section}{0pt}{1.5em}{0.6em}
\titlespacing*{\subsection}{0pt}{0.9em}{0.4em}

\fancypagestyle{plain}{
  \fancyhf{}
  \fancyfoot[R]{\footnotesize\thepage}
  
  }
\newcommand{\ket}[1]{\lvert #1\rangle}
\newcommand{\bra}[1]{\langle #1\rvert}
\newcommand{\braket}[2]{\langle #1\vert #2\rangle}
\newcommand{\Tr}{\mathrm{Tr}}
\newcommand{\Gmax}{G_{\max}}
\newcommand{\Gc}{G_{c}}
\newcommand{\Fphi}{F_{\phi\phi}}
\newcommand{\Icl}{\mathcal I}
\newcommand{\Dop}{\mathcal D}
\title{Gravitational acceleration encoded in Jaynes--Cummings exchange
frequencies: Quantum Fisher information, readout, and validity conditions}
\usepackage{orcidlink}
\author[1]{Salman Sajad Wani\,\orcidlink{0000-0002-5262-9738}
\thanks{Email: \href{mailto:sawa54922@hbku.edu.qa}
{\nolinkurl{sawa54922@hbku.edu.qa}}}}
\author[1]{Mughees Ahmed Khan\,\orcidlink{0009-0009-4214-7637}
\thanks{Email: \href{mailto:mukh68937@hbku.edu.qa}
{\nolinkurl{mukh68937@hbku.edu.qa}}}}
\author[2]{Mohammad Haris Khan\,\orcidlink{0009-0007-8927-484X}
\thanks{Email: \href{mailto:hariskhanhk9333@gmail.com}
{\nolinkurl{hariskhanhk9333@gmail.com}}}}
\author[2]{Fardeen Ahmad Sofi\,\orcidlink{0009-0004-0792-5353} 
\thanks{Email: \href{mailto:fardeenrafiq@gmail.com}
{\nolinkurl{fardeenrafiq@gmail.com}}}}
\author[3]{Abrar Ahmed Naqash\,\orcidlink{0000-0003-3891-4740}
\thanks{Corresponding author: 
\href{mailto:abrar.naqash@candqrc.ca}
{\nolinkurl{abrar.naqash@candqrc.ca}}}}
\author[1]{Saif Al-Kuwari\,\orcidlink{0000-0002-4402-7710}
\thanks{Email: \href{mailto:smalkuwari@hbku.edu.qa}
{\nolinkurl{smalkuwari@hbku.edu.qa}}}}
\affil[1]{Qatar Center for Quantum Computing, College of Science and Engineering,
Hamad Bin Khalifa University, Doha, Qatar}
\affil[2]{Department of Physics, University of Kashmir, Srinagar, 190006, India}
\affil[3]{Canadian Quantum Research Center, 204-3002 32 Ave, Vernon, BC V1T 2L7, Canada}
\date{}
\begin{document}
\maketitle
\thispagestyle{fancy}

\begin{abstract}
We derive an effective trapped atom--cavity model in which a constant
gravitational acceleration shifts the oscillator equilibrium and changes the
local standing-wave coupling, thereby encoding the acceleration in the
Jaynes--Cummings exchange frequencies. With the atom and motion initially in
their ground states and the cavity field initially coherent, we solve the
closed-system carrier dynamics exactly and derive the displaced-frame quantum
Fisher information (QFI) of the joint atom--cavity state. This QFI is
proportional to the square of the local coupling slope, grows quadratically
with interrogation time, and scales linearly with mean photon number. At the
node, phase-referenced Ramsey detection gives a sign-sensitive estimate of
axial acceleration and locally saturates the joint QFI. Away from the node,
photon counting and phase-optimized homodyne detection provide cavity readouts
when the cavity state carries more QFI than the atomic state. At the off-node
operating point studied, Lindblad simulations show that cavity loss produces a finite-time QFI optimum. Lamb Dicke and sideband-suppression conditions
control the carrier approximation. In the closed-system benchmark, the
carrier-model QFI agrees with the atom-cavity QFI obtained from the
unexpanded model after tracing out motion.
\end{abstract}

\vspace{0.6em}
\section{Introduction}
\label{sec:introduction}
Light-pulse atom gravimeters estimate gravitational acceleration through the
phase of freely falling matter waves.  Raman atom interferometers demonstrated this approach in precision measurements of
$g$ \cite{KasevichChu1992,PetersChungChu1999}, and atom interferometry provides
the broader matter-wave context for such inertial measurements \cite{
CroninSchmiedmayerPritchard2009}. In light-pulse Mach Zehnder geometries, the
leading acceleration-induced phase is proportional to $k_{\rm eff}gT^2$, where
$T$ is the free-evolution time between pulses and $k_{\rm eff}$ is the effective
momentum transfer \cite{PetersChungChu1999,AbendEtAl2016}. For the bound center-of-mass oscillator studied here, the encoding is different: gravity no longer appears as a free-fall propagation phase but as a static force on the harmonic coordinate. After the oscillator displacement, this force shifts the standing-wave sampling phase and thereby changes the Jaynes Cummings carrier coupling; the acceleration is therefore estimated through the corresponding carrier Rabi frequency. Within this {carrier-frequency encoding} model, we analyze the precision, {
atom-only and cavity-only readout channels}, and validity of the carrier-only
reduction. 

{The closest cavity-based comparisons encode $g$ through different dynamical variables and effective couplings. Cavity optomechanics couples optical and mechanical degrees of freedom through radiation pressure \cite{KippenbergVahala2007}; in nonlinear optomechanical gravimetry, this coupling maps the gravitationally driven mechanical dynamics into the phase of the
optical output, which can be read out by homodyne detection \cite{QvarfortEtAl2018}. Cold-atom cavity systems realize dispersive and collective light-matter couplings \cite{RitschDomokosBrenneckeEsslinger2013}; in a gravimetric sensor based on a Bose--Einstein condensate in a double well, gravity modifies effective light-matter parameters and is inferred from atomic or photonic measurements \cite{NiezgodaEtAl2021}.These schemes rely on radiation-pressure or dispersive and collective couplings rather than the transverse carrier exchange considered here.} 

Jaynes Cummings physics has also appeared in force-sensing contexts.A two-photon Raman study showed that a gravity-induced displacement from a standing-wave node produces a residual carrier and reduces its suppression relative to the motional sidebands \cite{ReimannEtAl2014}.
 In trapped ions, Jaynes Cummings, quantum-Rabi, and Jahn Teller spin--motion models have been used to transfer weak-force information to Ramsey oscillations of an internal spin \cite{IvanovVitanovSinger2016}. Homogeneous gravity has likewise been studied for moving atoms in a traveling-wave Jaynes--Cummings setting, where it modifies atom--field dynamics, momentum diffusion, photon statistics, and field quadratures rather than encoding a carrier Rabi frequency in a trapped standing-wave system\cite{MohammadiNaderiSoltanolkotabi2007}.In this paper, we study the resonant Jaynes--Cummings dynamics of a single atom coupled to a quantized standing-wave cavity mode, with the gravity-induced equilibrium shift encoded in the exchange frequencies. We derive the joint QFI, evaluate the atomic and cavity reduced-state QFIs, identify the corresponding local readouts, and establish the regime in which the carrier reduction remains controlled.

{Within the optical rotating-wave approximation (RWA), we start from the standard trapped cavity-QED setting of an internal atomic transition coupled to a cavity mode while the center-of-mass motion is quantized \cite{
BuzekEtAl1997,ZippilliMorigi2005}. We specialize this setting to a single
standing-wave cavity mode and one center-of-mass oscillator along the cavity
axis.}  The gravitational acceleration enters the mechanical Hamiltonian as a constant force. An exact oscillator displacement eliminates the corresponding linear term and recenters the oscillator at its shifted equilibrium. The gravitational dependence remains in the atom--cavity coupling through the resulting shift of the local standing-wave phase. { In the Lamb Dicke and
resolved-sideband regimes, this shifted sampling phase sets the retained
phonon-preserving carrier coupling in a trapped-particle cavity-QED
reduction \cite{WuYang1997,ZippilliMorigi2005,
LeibfriedBlattMonroeWineland2003,WinelandMonroeItano1998}. The retained
transverse exchange has the standard Jaynes Cummings form \cite{
JaynesCummings1963,ShoreKnight1993}, while higher-order Lamb Dicke terms enter as corrections to the carrier-only reduction \cite{
LeibfriedBlattMonroeWineland2003,WinelandMonroeItano1998}.} The resulting
carrier coupling sets the exchange frequencies of the Jaynes Cummings ladder,
thereby encoding the static acceleration as a Rabi frequency.

We solve the closed dynamics exactly for the motional ground state, the atomic internal ground state, and a cavity field initially prepared in a coherent state. Conservation of excitation number reduces the atom-cavity
evolution to independent two-dimensional Jaynes Cummings blocks, and a coherent
probe populates a ladder of Rabi frequencies with the usual $\sqrt n$
dependence. The exact state { determines} the atomic signal, the
cavity and atomic reduced density operators, and the atom-cavity entanglement
entropy. {the atomic and cavity reductions determine how this information is distributed between the atom-only and cavity-only
states: near a standing-wave node, the regular, sign-sensitive readout is a
transverse atomic pseudospin component measured with a Ramsey analysis pulse,
whereas at off-node operating points where the cavity reduced state carries more quantum Fisher information (QFI) than the atomic state, photon counting and phase-optimized homodyne detection provide cavity readouts.} For the atom-cavity sensing protocol in the displaced oscillator frame, we compute the pure-state QFI of the exact atom-cavity state \cite{BraunsteinCaves1994,Paris2009}.The QFI grows quadratically with interrogation time, scales linearly with the coherent state's mean photon number, and is controlled by the square of the standing-wave slope at the operating point. { We distinguish this
displaced-frame atom-cavity QFI from the lab-frame product-state QFI, which
contains an additional mechanical displacement contribution.} 

The standing-wave slope that produces gravitational
responsivity also generates phonon-assisted sidebands. Thus the controlled
carrier-only limit requires both a Lamb Dicke expansion and resolved-sideband
suppression. We quantify these requirements using a Lamb--Dicke support parameter and first- and second-order sideband diagnostics, $\epsilon_{\rm LD}$, $\epsilon_{1}$, and $\epsilon_{2}$, all of which must remain small for the carrier-only Hamiltonian to provide a controlled effective description.  Finally, we include cavity loss through a Lindblad master equation \cite{Lindblad1976,GoriniKossakowskiSudarshan1976}. At the off-node point \(\phi=\pi/4\), simulations show that cavity loss replaces the ideal quadratic-in-time QFI growth with a maximum at a finite interrogation time.
The remainder of the paper follows this sequence:
Sec.~\ref{sec:model} develops the model and validity conditions,
Secs.~\ref{sec:dynamics} and \ref{sec:qfi_readout} present the closed-system
dynamics and metrology, and Sec.~\ref{sec:open} gives the open-system analysis.

\section{Derivation of the gravity-biased carrier--Jaynes Cummings Hamiltonian}
\label{sec:model}
{The effective carrier--Jaynes Cummings Hamiltonian follows from an
optical-RWA laboratory-frame cavity-QED model with a static force $mg\hat x$.
Displacing the forced oscillator moves the acceleration dependence into the
standing-wave sampling phase: the atom samples the cavity mode at the shifted
phase $\theta_g=kx_0-kg/\omega_m^2$. The subsequent Lamb Dicke and
resolved-sideband reduction retains only the phonon-preserving carrier exchange
and omits the phonon-assisted sidebands, yielding the carrier-only model analyzed below.
}

\subsection{Laboratory model and static-force displacement}

We consider a single two-level atom of transition frequency $\omega_q$, a single
cavity mode of frequency $\omega_c$, and one quantized COM mode of frequency
$\omega_m$ along the cavity axis. { The estimated acceleration is
denoted by lowercase $g$, while capital $G$ denotes atom-cavity couplings such
as $G(x)$, $\Gmax$, $G_0$, $G_1$, and $\Gc$. The standing-wave phase of the trap
center is $\theta_0=kx_0$.} The position operator is
\begin{equation}
        \hat x=x_0+x_{\rm zpf}(b+b^\dagger),
        \qquad
        x_{\rm zpf}=\sqrt{\frac{\hbar}{2m\omega_m}},
\end{equation}
where $x_0$ is the trap center measured relative to the standing-wave mode
pattern. Before the optical rotating-wave approximation, the
single-mode dipole interaction is taken in the Rabi form
\begin{equation}
{
        H_{\rm int}^{\rm Rabi}
        =
        \hbar G(\hat x)(a+a^\dagger)(\sigma_++\sigma_-).
}
\end{equation}
The product expansion separates the near-resonant exchange terms
$a\sigma_+$ and $a^\dagger\sigma_-$ from the counter-rotating terms
$a\sigma_-$ and $a^\dagger\sigma_+$. In the interaction picture, the
former rotate at the detuning scale $\Delta=\omega_q-\omega_c$, whereas the
latter rotate at frequencies of order $\omega_q+\omega_c$; the explicit
bookkeeping is given in Appendix\ref{app:LD},
Eq.\eqref{eq:app_optical_rwa_terms}. For
$|\Delta|=|\omega_q-\omega_c|\ll\omega_q+\omega_c$ and light-matter matrix elements small compared with the optical frequencies, the counter-rotating terms average out on the resonant timescale \cite{JamesJerke2007}. The retained near-resonant interaction is
\begin{equation}
{
        H_{\rm int}^{\rm RWA}
        =
        \hbar G(\hat x)(a\sigma_+ + a^\dagger\sigma_-).
}
\label{eq:Hint_RWA}
\end{equation}
With gravity along the cavity-axis COM coordinate, the optical-RWA
laboratory-frame Hamiltonian obtained from Eq.\eqref{eq:Hint_RWA} is
\begin{equation}
\hat H_{\rm lab}(g)=
\hbar\omega_c a^\dagger a
+\frac{\hbar\omega_q}{2}\sigma_z
+\hbar\omega_m b^\dagger b
+mg\hat x
+\hbar G(\hat x)(a\sigma_+ + a^\dagger\sigma_-),
\label{eq:Hlab}
\end{equation}
with $G(\hat x)=\Gmax\cos(k\hat x)$. The only explicit dependence on $g$ is
the linear mechanical potential $mg\hat x$, corresponding to a constant force on
the oscillator. The mechanical-plus-gravity sector is
\begin{equation}
        \hat H_m(g)=
        \hbar\omega_m b^\dagger b
        +mgx_0
        +mgx_{\rm zpf}(b+b^\dagger).
\end{equation}
The force-canceling displacement is parametrized by
\begin{equation}
        \beta_g=\frac{mgx_{\rm zpf}}{\hbar\omega_m},
        \qquad
        \widetilde D(\beta_g)=\exp[-\beta_g(b^\dagger-b)].
\label{eq:beta_D_main}
\end{equation}
The displacement rules are
\begin{equation}
        \widetilde D^\dagger b\widetilde D=b-\beta_g,
        \qquad
        \widetilde D^\dagger b^\dagger\widetilde D=b^\dagger-\beta_g.
\end{equation}
These rules shift the oscillator coordinate by
$-2x_{\rm zpf}\beta_g$. With the choice in Eq.\eqref{eq:beta_D_main}, the
coefficient of $b+b^\dagger$ cancels exactly, so the transformed mechanical
Hamiltonian is a bare oscillator plus a scalar phase. The BCH derivation and the
cancellation algebra are given in Appendix\ref{app:disp},
Eqs.\eqref{eq:app_disp_rules}--\eqref{eq:app_xeq_derivation}. The resulting
equilibrium shift is
\begin{equation}
        x_{\rm eq}(g)=-2x_{\rm zpf}\beta_g=-\frac{g}{\omega_m^2}.
\end{equation}

%%%%%%%%%%%%%%%%%%%%%%%%%%%%%%%%%%%%%%%%%%%%%%%%%%%%%%%%%%%%%%%%%%%%%%%%
%%%%%%%%%%%%%%%%%%%%%%%%%%%%%%%%%%%%%%%%%%%%%%%%%%%%%%%%%%%%%%%%%%%%%
\subsection{Gravity-biased carrier coupling}

The displacement transfers the $g$-dependence from the mechanical force term
to the phase at which the atom samples the standing wave. In the displaced
frame,
\begin{equation}
G(\hat x)=\Gmax\cos\!\left[
        kx_0+kx_{\rm zpf}(b+b^\dagger)-\frac{kg}{\omega_m^2}
        \right].
\label{eq:G_displaced}
\end{equation}
With $\theta_0=kx_0$, define the gravity-induced standing-wave phase shift and
the shifted sampling phase by
$\phi=kg/\omega_m^2$ and $\theta_g=\theta_0-\phi$.
With $\eta=kx_{\rm zpf}$ and $X=b+b^\dagger$,
Eq.~\eqref{eq:G_displaced} becomes
$G(\hat x)=\Gmax\cos(\theta_g+\eta X)$, so $g$ enters the coupling only
through the shifted phase $\theta_g$.

The Lamb Dicke expansion requires the dimensionless motional displacement
$\eta X$ to remain small over the occupied motional sector. For a motional
state with number probabilities $p_m$, let $m_{\rm eff}$ denote a
high-probability cutoff satisfying
$\sum_{m=0}^{m_{\rm eff}}p_m\ge 1-\varepsilon_m .$
For each number state in the retained support, a sufficient Lamb Dicke
condition is
\begin{equation}
        \eta\sqrt{\langle m|X^2|m\rangle}\ll1,
        \qquad
        \langle m|X^2|m\rangle=2m+1,
        \qquad
        \eta\sqrt{2m_{\rm eff}+1}\ll1 .
\label{eq:LD_support_condition}
\end{equation}
For the displaced motional ground-state protocol, $m_{\rm eff}=0$, and
Eq.~\eqref{eq:LD_support_condition} reduces to $\eta\ll1$. Expanding
Eq.~\eqref{eq:G_displaced} to first order in $\eta$ gives
\begin{equation}
\begin{aligned}
        G(\hat x)
        &=
        \Gmax\cos[\theta_g+\eta(b+b^\dagger)]
        \\
        &=
        G_0(g)+G_1(g)(b+b^\dagger)+O(\eta^2),
        \\
        G_0(g)
        &=
        \Gmax\cos\theta_g,
        \qquad
        G_1(g)
        =
        -\Gmax\eta\sin\theta_g .
\end{aligned}
\label{eq:LD_main}
\end{equation}
Retaining the phonon-number-preserving term in
Eq.~\eqref{eq:LD_main} gives the leading carrier coupling
\begin{equation}
        \Gc(g)\equiv G_0(g)
        =
        \Gmax\cos\!\left(kx_0-\frac{kg}{\omega_m^2}\right).
\label{eq:Gc_LD}
\end{equation}
In the resonant Jaynes Cummings interaction, $\Gc(g)$ sets the exchange
frequencies of the $n$-excitation manifolds,
$\Omega_n(g)=\Gc(g)\sqrt n .$
Hence the acceleration dependence of $\Omega_n(g)$ enters through the
shifted phase $\theta_g$ in $\Gc(g)$. Differentiating
Eq.~\eqref{eq:Gc_LD} gives the gravitational responsivity
\begin{equation}
\partial_g\Gc(g)=
\Gmax\frac{k}{\omega_m^2}
\sin\!\left(kx_0-\frac{kg}{\omega_m^2}\right).
\label{eq:dGc}
\end{equation}
Comparison with the sideband coefficient $G_1(g)$ in
Eq.~\eqref{eq:LD_main} gives
\begin{equation}
        \partial_g\Gc(g)
        =
        -\frac{G_1(g)}{x_{\rm zpf}\omega_m^2}.
\label{eq:responsivity_sideband_relation}
\end{equation}
Thus the same standing-wave slope that produces gravitational responsivity also
sets the amplitude of the leading phonon-assisted sideband.

At node bias $\theta_0=\pi/2$, where the zero-acceleration position
coincides with a standing-wave node,
\begin{equation}
        \Gc(g)
        =
        \Gmax\sin\!\left(\frac{kg}{\omega_m^2}\right)
        \simeq
        \Gmax\frac{kg}{\omega_m^2},
        \qquad
        \left|\frac{kg}{\omega_m^2}\right|\ll1 .
\end{equation}
At $g=0$, $\Gc$ vanishes while $|\partial_g\Gc|$ is maximal; for
$|kg/\omega_m^2|\ll1$, the carrier is linear in the axial acceleration.

\begin{figure}[tbp]
  \centering
  \includegraphics[width=\linewidth]{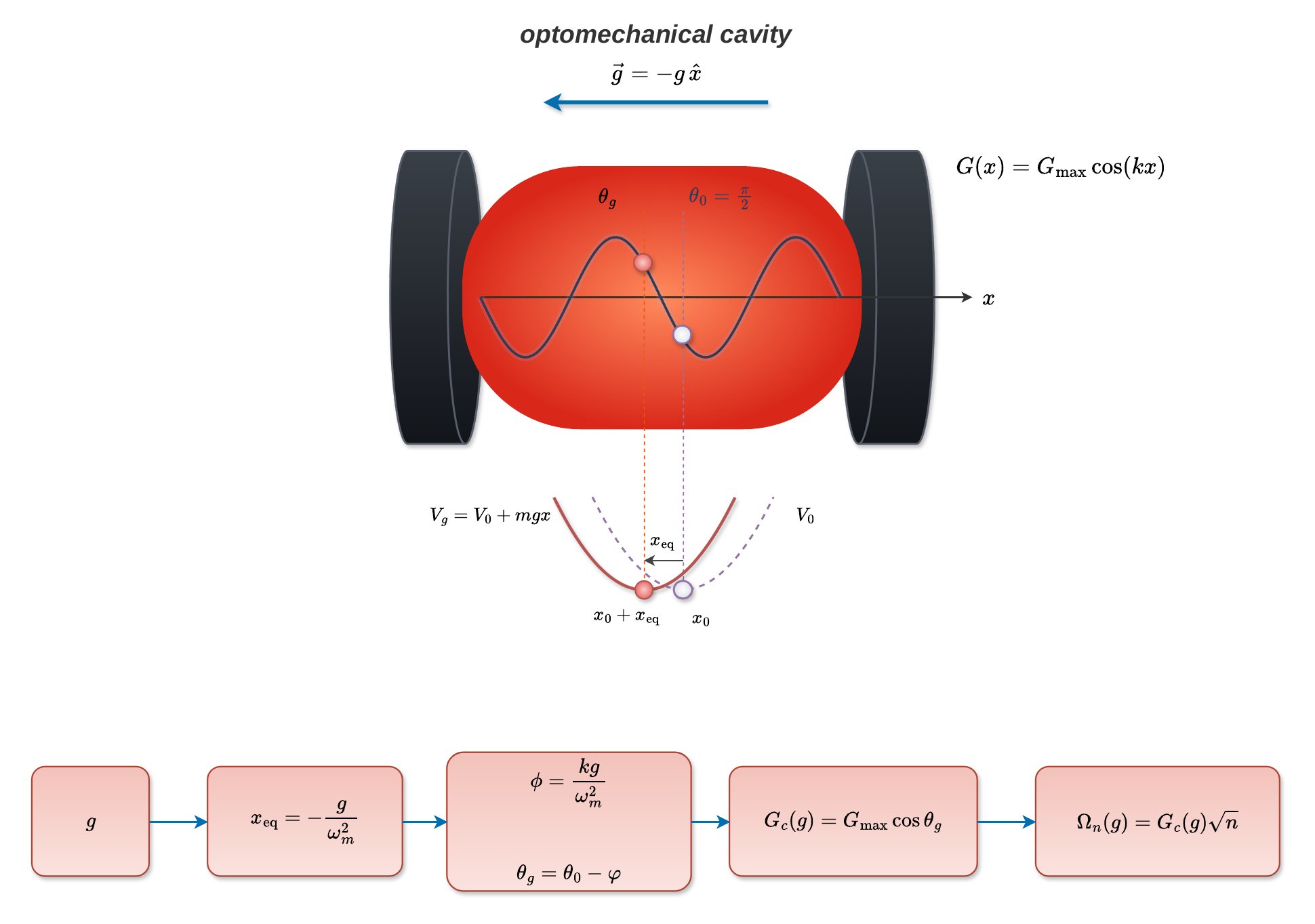}
  \caption{\textbf{Gravity-to-carrier-frequency transduction.}
  A gravitational acceleration component along the cavity axis appears as a
  static force on the bound center-of-mass oscillator and shifts its equilibrium
  from $x_0$ to $x_0+x_{\rm eq}$, with $x_{\rm eq}=-g/\omega_m^2$. The
  equilibrium shift changes the standing-wave phase sampled by the atom to
  $\theta_g=\theta_0-\phi$, where $\phi=kg/\omega_m^2$. Retaining the leading
  Lamb Dicke carrier gives $\Gc(g)=\Gmax\cos\theta_g$, and the Jaynes Cummings
  ladder converts this coupling into the exchange frequencies
  $\Omega_n(g)=\Gc(g)\sqrt n$. Filled points mark positions on the cavity axis,
  while open circles mark the corresponding values of the standing-wave coupling
  profile. The displacement is schematic and not to scale.}
  \label{fig:transduction}
\end{figure}

The gravity-to-carrier-frequency transduction chain,
$g\rightarrow x_{\rm eq}(g)\rightarrow\theta_g\rightarrow
\Gc(g)\rightarrow\Omega_n(g)$, is shown schematically in
Fig.~\ref{fig:transduction}.

\subsection{Carrier-JC reduction and validity diagnostics}

{
For a motional number state $\ket m$, the exact diagonal coupling matrix
element is
$\langle m|G(\hat x)|m\rangle
=\Gmax e^{-\eta^2/2}L_m(\eta^2)\cos\theta_g$
(Appendix~\ref{app:LD}). For $m=0$, it becomes
$\Gmax e^{-\eta^2/2}\cos\theta_g$, differing from
Eq.~\eqref{eq:Gc_LD} only at $O(\eta^2)$. The Debye--Waller factor
$e^{-\eta^2/2}$ rescales both the ground-state carrier and its gravitational
responsivity and is omitted in the leading-order carrier approximation.

To bound the sideband matrix elements for a coherent probe with mean photon
number $\bar n$, define $n_{\rm eff}$ as the smallest integer for which
the Poisson tail does not exceed $\varepsilon_n$ \cite{VolkoffSubasi2022}:
\begin{equation}
        n_{\rm eff}(\bar n,\varepsilon_n)
        =
        \min\left\{
        N\in\mathbb N_0:
        1-\sum_{n=0}^{N}
        e^{-\bar n}\frac{\bar n^n}{n!}
        \le \varepsilon_n
        \right\}.
\label{eq:neff_def}
\end{equation}
The sectors $0\le n\le n_{\rm eff}$ and
$0\le m\le m_{\rm eff}$ therefore contain probabilities of at least
$1-\varepsilon_n$ and $1-\varepsilon_m$, respectively. These cutoffs
delimit the photon- and phonon-number sectors over which the sideband matrix
elements are bounded at the prescribed tail tolerances.
}

The phonon-assisted term $G_1(b+b^\dagger)$ is rapidly rotating in the
mechanical interaction picture. On atom-cavity resonance, $\omega_q=\omega_c$,
the carrier term is time independent, while the sideband term is
\begin{equation}
V_{\rm sb}^{(I)}(t)=
\hbar G_1(g)
\left(be^{-i\omega_m t}+b^\dagger e^{i\omega_m t}\right)
(a\sigma_+ + a^\dagger\sigma_-).
\label{eq:Vsb_main}
\end{equation}
The first-order sideband contribution is bounded because its propagator contains
the oscillatory integral
$\int_0^t ds\,e^{\pm i\omega_m s}=O(\omega_m^{-1})$. The relevant small
parameter is therefore the sideband matrix element, evaluated over the retained
photon and phonon support, divided by $\omega_m$.
Thus the sideband is perturbatively suppressed when its largest retained
matrix element is small compared with $\omega_m$,
\begin{equation}
        \omega_m\gg
        |G_1(g)|\sqrt{(n_{\rm eff}+1)(m_{\rm eff}+1)} .
\label{eq:rsb}
\end{equation}
Here $n_{\rm eff}$ and $m_{\rm eff}$ are defined by explicit photon and
phonon tail tolerances. For the displaced motional ground-state protocol,
$m_{\rm eff}=0$, and Eq.~\eqref{eq:rsb} reduces to
$\omega_m\gg |G_1|\sqrt{n_{\rm eff}+1}$. The numerical value of this bound
depends on the specified $\varepsilon_n$, $\varepsilon_m$,
$n_{\rm eff}$, $m_{\rm eff}$, and Fock cutoff.
For the effective open-system model, the same scale separation requires
$\omega_m\gg\kappa_{\rm ph},\gamma_{\rm ph},\gamma_{\phi,{\rm ph}}$, or
equivalently
$\omega_m/\Gmax\gg\tilde\kappa,\tilde\gamma,\tilde\gamma_\phi$, so that the
dissipative rates remain small compared with the sideband detuning scale
$\omega_m$.
When Eq.~\eqref{eq:rsb} holds, the leading effective Hamiltonian is the resonant
transverse JC model
\begin{equation}
\hat H_{\rm eff}(g)=
\hbar\omega_c a^\dagger a
+\frac{\hbar\omega_q}{2}\sigma_z
+\hbar\Gc(g)(a\sigma_+ + a^\dagger\sigma_-),
\qquad \omega_q=\omega_c.
\label{eq:Heff}
\end{equation}
The leading averaged correction generated by the first-order sideband term
enters at order $O(G_1^2/\omega_m)$ in a Magnus expansion
\cite{GoldmanDalibard2014}.
The commutator estimate is given in Appendix~\ref{app:LD},
Eqs.~\eqref{eq:app_Hpm}--\eqref{eq:app_Hsb_scale}. If this averaged correction
is omitted, its accumulated phase is tracked by the diagnostic
$\epsilon_2$ below.
The validity of the carrier-only model is tracked by the dimensionless
quantities $\epsilon_{\rm LD}$, $\epsilon_1$, and $\epsilon_2$, whose
numerical values require the physical scales $\omega_m$ and $\Gmax$, the
photon and motional supports, and the loss rates. With the support definitions
in Eqs.~\eqref{eq:neff_def} and \eqref{eq:LD_support_condition}, these
quantities are
\begin{equation}
\begin{aligned}
        \epsilon_{\rm LD}
        &=
        \eta\sqrt{2m_{\rm eff}+1},
        \\
        \epsilon_1
        &=
        \frac{|G_1|}{\omega_m}
        \sqrt{(n_{\rm eff}+1)(m_{\rm eff}+1)},
        \\
        \epsilon_2
        &=
        t\,\frac{G_1^2}{\omega_m}n_{\rm eff}
        =
        \tau\,\frac{G_1^2}{\Gmax\omega_m}n_{\rm eff}.
\end{aligned}
\end{equation}

{
For $m_{\rm eff}=0$, these expressions reduce to the
motional-ground-state limits. The carrier-only model requires
$\epsilon_{\rm LD}\ll1$, $\epsilon_1\ll1$, and $\epsilon_2\ll1$.}

{
\subsection{Benchmark against the unexpanded motional model}

A representative closed-system benchmark within this regime uses
$\omega_m/\Gmax=10$, $\eta=9.71\times10^{-3}$,
$\theta_0=\pi/2$, and $\bar n=9$, with $m_{\rm eff}=0$.
For the coherent-state tail tolerance $\varepsilon_n=10^{-10}$,
Eq.~\eqref{eq:neff_def} gives $n_{\rm eff}=34$. Both evolutions start from
$ \ket{\Psi(0)}=\ket{0_m}\otimes\ket g\otimes\ket\alpha,$
       with $ |\alpha|^2=\bar n .$
At the maximal standing-wave slope, $|\sin\theta_g|=1$, and the maximum
interrogation time, $\tau_{\max}=12\pi$, the diagnostics are
$\epsilon_{\rm LD}=9.71\times10^{-3}$,
$\epsilon_1=5.75\times10^{-3}$, and
$\epsilon_2=1.21\times10^{-2}$. The unexpanded displaced-frame Hamiltonian used for comparison is
\begin{equation}
        \frac{H_{\rm unexp}(\phi)}{\hbar\Gmax}
        =
        \frac{\omega_m}{\Gmax}b^\dagger b
        +
        \cos\!\left[\theta_g+\eta(b+b^\dagger)\right]
        (a\sigma_+ + a^\dagger\sigma_-).
\label{eq:Hunexp_validation}
\end{equation}
It retains the complete standing-wave dependence while preserving the optical
RWA and exact atom--cavity resonance assumed in the carrier model. At
dimensionless time $\tau$, its state is
\begin{equation}
        \ket{\Psi_{\rm unexp}(\tau;\phi)}
        =
        \exp\!\left[
        -i\tau\frac{H_{\rm unexp}(\phi)}{\hbar\Gmax}
        \right]
        \ket{\Psi(0)} .
\label{eq:Psi_unexp_validation}
\end{equation}
The motional mode is traced out before the atom--cavity QFI is evaluated:
\begin{equation}
        \rho_{ac}^{\rm unexp}(\tau;\phi)
        =
        \Tr_m\!\left[
        \ket{\Psi_{\rm unexp}(\tau;\phi)}
        \bra{\Psi_{\rm unexp}(\tau;\phi)}
        \right].
\label{eq:rhoac_unexp_validation}
\end{equation}
The two atom--cavity QFIs are therefore defined on the same subsystem. The relative atom--cavity-QFI discrepancy and the absolute difference between
atom-only QFI fractions are
\begin{align}
\varepsilon_{ac}(\tau)
&=
\frac{
\left|
F_{\phi\phi}^{ac,\mathrm{unexp}}(\tau)
-
F_{\phi\phi}^{ac,\mathrm{eff}}(\tau)
\right|
}{
F_{\phi\phi}^{ac,\mathrm{unexp}}(\tau)
},
\label{eq:validation_error_ac}
\\
\Delta f_{\rm at}(\tau)
&=
\left|
\frac{
F_{\phi\phi}^{\rm at,\mathrm{unexp}}(\tau)
}{
F_{\phi\phi}^{ac,\mathrm{unexp}}(\tau)
}
-
\frac{
F_{\phi\phi}^{\rm at,\mathrm{eff}}(\tau)
}{
F_{\phi\phi}^{ac,\mathrm{eff}}(\tau)
}
\right|.
\label{eq:validation_error_at}
\end{align}
These quantities are evaluated at times for which both atom--cavity QFIs are
nonzero.
Figure~\ref{fig:full_model_validation} presents the motion-traced atom--cavity
QFI and atom-only QFI fraction for both models at the node and at the
representative off-node point $\phi=\pi/4$.
\begin{figure}[t]
\centering
\includegraphics[width=\textwidth]{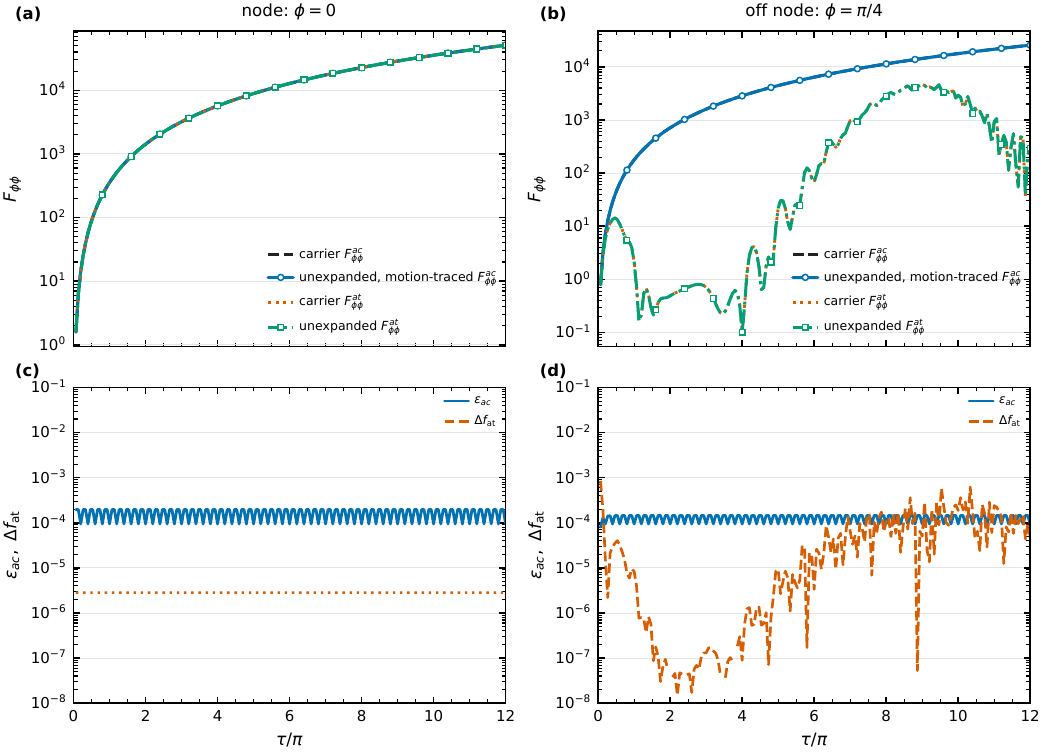}
\caption{\textbf{Validation of the carrier-only reduction.}
Panels (a) and (c) show the node operating point, $\phi=0$, whereas panels
(b) and (d) show the off-node point $\phi=\pi/4$. The upper panels compare
the atom--cavity and atom-only QFIs of the effective carrier model with those
obtained from the unexpanded displaced-frame Hamiltonian in
Eq.~\eqref{eq:Hunexp_validation}. For the unexpanded model, the motional mode
is traced out before evaluating the atom--cavity QFI, as defined in
Eq.~\eqref{eq:rhoac_unexp_validation}. The lower panels show
$\varepsilon_{ac}$ and $\Delta f_{\rm at}$ from
Eqs.~\eqref{eq:validation_error_ac} and
\eqref{eq:validation_error_at}. For the benchmark specified above, the maximum
relative atom--cavity-QFI discrepancies over the sampled nonzero times
$0<\tau\le12\pi$ are $1.96\times10^{-4}$ at the node and
$1.45\times10^{-4}$ at $\phi=\pi/4$.}
\label{fig:full_model_validation}
\end{figure}
Convergence and tangent-state checks are given in
Appendix~\ref{app:numerics}. For this benchmark, the carrier model reproduces
the motion-traced atom--cavity QFI of the unexpanded model over
$0<\tau\le12\pi$. The dissipative analysis in
Sec.~\ref{sec:open} uses the effective carrier model.
}

%%%%%%%%%%%%%%%%%%%%%%%%%%%%%%%%%%%%%%%%%%%%%%
\section{Closed-system dynamics and reduced observables}
\label{sec:dynamics}
%%%%%%%%%%%%%%%%%%%%%%%%%%%%%%%%%%%%%%%%%%%%%%%%%%%%%%%%%%%%%%%%%%%%%

\subsection{Exact Jaynes Cummings state}

The effective Hamiltonian acts only on the atom-cavity
Hilbert factor, $\mathcal H_{\rm at}\otimes\mathcal H_{\rm cav}$, so the
displaced-frame motional ground state $\ket{0_m}$
factorizes throughout the closed dynamics.
The closed-system evolution starts from
\begin{equation}
        \ket{\Psi(0)}=
        \ket{0_m}\otimes\ket g\otimes\ket\alpha,
        \qquad
        \ket\alpha=e^{-|\alpha|^2/2}\sum_{n=0}^\infty
        \frac{\alpha^n}{\sqrt{n!}}\ket n,
\end{equation}
with $c_n=e^{-|\alpha|^2/2}\alpha^n/\sqrt{n!}$ and mean photon number
$\bar n=|\alpha|^2$.
Conservation of the total excitation number
$N=a^\dagger a+\sigma_+\sigma_-$ under the exchange operator
$K=a\sigma_+ + a^\dagger\sigma_-$ decomposes the dynamics into independent
fixed-excitation JC blocks. In the $N=n$
block,
$\{\ket{g,n},\ket{e,n-1}\}$, the exchange acts as $\sqrt n\,\tau_x$. The
block-diagonal derivation and propagator action are given in
Appendix\ref{app:prop}, Eqs.\eqref{eq:app_JC_conservation}--\eqref{eq:app_UI_action}.
The resulting $n$-dependent Rabi frequency is
\begin{equation}
        \Omega_n(g)=\Gc(g)\sqrt n.
\end{equation}
On exact atom-cavity resonance, $\omega_q=\omega_c\equiv\omega$, the full
atom-cavity state is
\begin{equation}
\begin{aligned}
\ket{\psi(t;g)}
=&\;c_0e^{i\omega t/2}\ket{g,0}
\\
&+\sum_{n=1}^\infty
c_ne^{-i\omega(n-\frac12)t}
\left[
\cos(\Omega_n t)\ket{g,n}
-i\sin(\Omega_n t)\ket{e,n-1}
\right].
\end{aligned}
\label{eq:psi_full_phase}
\end{equation}
{
Removing the common free phase factor in each fixed-excitation block,
for example $e^{-i\omega(n-\frac12)t}$ in the $N=n$
block, and defining
$\tau=\Gmax t$ and $u(\phi)=\Gc(g)/\Gmax=\cos(\theta_0-\phi)$ gives the
interaction-picture state}
\begin{equation}
\ket{\psi(\tau;\phi)}=
 c_0\ket{g,0}
 +\sum_{n=1}^\infty c_n
\left[
\cos(\sqrt n\,u\tau)\ket{g,n}
-i\sin(\sqrt n\,u\tau)\ket{e,n-1}
\right],
\end{equation}
Gravity appears only through the Rabi phases $\sqrt n\,u\tau$.
The atomic excited-state probability is
\begin{equation}
        P_e(\tau;\phi)=
        \sum_{n=1}^\infty |c_n|^2\sin^2(\sqrt n\,u\tau).
\label{eq:Pe}
\end{equation}

\begin{figure}[t]
\centering
\includegraphics[width=\textwidth]{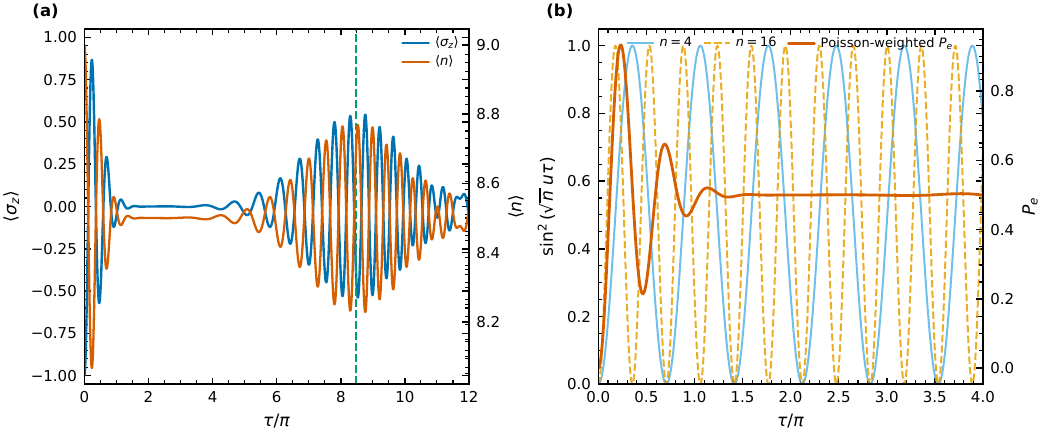}
\caption{\textbf{Closed coherent oscillations} at
$\theta_0=\pi/2$, $\phi=\pi/4$, and $\bar n=9$.
(a) Atomic inversion $\langle\sigma_z\rangle$ and mean photon number
$\langle n\rangle$ exchange energy coherently. The $\sqrt n$ frequency spread
causes collapse and revival near
$\tau_{\rm rev}\approx2\pi\sqrt{\bar n}/u\approx8.5\pi$.
(b) Representative fixed-excitation blocks, $n=4$ and
$n=16$, oscillate at $\Omega_n\propto\sqrt n$; the Poisson-weighted sum gives
the collapsed and revived atomic signal $P_e(\tau)$.}
\label{fig:osc}
\end{figure}

A coherent cavity-mode state populates many photon-number components.
Each component enters a fixed-excitation JC block spanned by
$\{\ket{g,n},\ket{e,n-1}\}$, where the exchange frequency is
$\Omega_n=\Gc\sqrt n$. The population signal collapses as these
$\sqrt n$-dependent components dephase and revives when adjacent components
rephase \cite{RempeWaltherKlein1987}. For a coherent-state
photon distribution with Poisson width
small compared with its mean, $\sqrt{\bar n}/\bar n\ll1$ equivalently
$\bar n\gg1$, the revival estimate follows by setting the
phase difference between adjacent $n$ and $n+1$
frequency components, $2u\tau(\sqrt{n+1}-\sqrt n)$, to $2\pi$,
evaluating the
spacing near $n\simeq\bar n$, and using
$\sqrt{n+1}-\sqrt n\simeq1/(2\sqrt{\bar n})$. The corresponding asymptotic
revival time is
\begin{equation}
{
        \tau_{\rm rev}\approx \frac{2\pi\sqrt{\bar n}}{|u|}.
}
\label{eq:revival}
\end{equation}
For the numerical example $\theta_0=\pi/2$, $\phi=\pi/4$, and $\bar n=9$, this
gives $\tau_{\rm rev}\approx 8.5\pi$.
{
Figure\ref{fig:osc} plots the exact JC dynamics. Panel (a) shows the collapse
and revival of the atomic inversion together with the corresponding mean photon
number. Panel (b) shows the $n$-resolved JC-block oscillations
at
$\Omega_n=\Gc\sqrt n$, whose Poisson-weighted sum gives $P_e(\tau)$. The
plotted curves are direct evaluations of Eq.\eqref{eq:Pe}; the revival marker
uses Eq.\eqref{eq:revival}.
}

%%%%%%%%%%%%%%%%%%%%%%%%%%%%%%%%%%%%%%%%%%%%%%%%%%%%%%%%%%%%%%%%%%%%%%%%%%%%%%%%%%%%%%%%%%%%%

\subsection{Reduced states and entanglement entropy}

Although the global atom-cavity state is pure, most measurements access
either the atom or the cavity alone. The state in
Eq.\eqref{eq:psi_full_phase} can be decomposed into
atomic conditional cavity states as
\begin{equation}
        \ket{\psi(t;g)}=
        \ket g\otimes\ket{\phi_g(t;g)}+
        \ket e\otimes\ket{\phi_e(t;g)},
\label{eq:conditional_form}
\end{equation}
where
$\ket{\phi_g}=\sum_{n=0}^\infty A_n\ket n$ and
$\ket{\phi_e}=\sum_{n=0}^\infty E_n\ket n$, with coefficients
\begin{equation}
        A_n=c_ne^{-i\omega(n-\frac12)t}\cos(\Omega_n t),
        \qquad
        E_n=-ic_{n+1}e^{-i\omega(n+\frac12)t}\sin(\Omega_{n+1}t).
\end{equation}
Here $\Omega_0=0$. Tracing over the atom gives
$\rho_{\rm cav}=\ket{\phi_g}\bra{\phi_g}+\ket{\phi_e}\bra{\phi_e}$. In the
Fock basis,
\begin{equation}
\rho_{nn'}=
 e^{-i\omega(n-n')t}
\left[
 c_nc_{n'}^*\cos(\Omega_n t)\cos(\Omega_{n'}t)
 +c_{n+1}c_{n'+1}^*\sin(\Omega_{n+1}t)\sin(\Omega_{n'+1}t)
\right].
\label{eq:rho_cav_elements}
\end{equation}
Tracing over the cavity gives the atomic qubit
\begin{equation}
\rho_{\rm at}=
\begin{pmatrix}
\rho_{ee} & \rho_{eg}\\
\rho_{ge} & \rho_{gg}
\end{pmatrix}_{\{\ket e,\ket g\}},
\label{eq:rho_at}
\end{equation}
with
\begin{align}
\rho_{ee}
&=\sum_{n=0}^\infty |c_{n+1}|^2\sin^2(\Omega_{n+1}t),
\label{eq:rhoee}
\\
\rho_{gg}
&=\sum_{n=0}^\infty |c_n|^2\cos^2(\Omega_n t),
\\
\rho_{eg}
&=-ie^{-i\omega t}\sum_{n=0}^\infty
c_{n+1}c_n^*\sin(\Omega_{n+1}t)\cos(\Omega_n t),
\qquad
\rho_{ge}=\rho_{eg}^*.
\label{eq:rhoeg}
\end{align}
Population readout, i.e.\ measurement in the
$\{\ket e,\ket g\}$ basis, accesses $\rho_{ee}=P_e$, whereas Ramsey-type
atomic readout measures a transverse atomic quadrature sensitive to the
coherence $\rho_{eg}$.

For the pure atom-cavity state, $\rho_{\rm at}$ and $\rho_{\rm cav}$ have the
same nonzero spectra. Because the atomic subsystem is two dimensional, the
Schmidt rank is at most two and the entanglement entropy obeys
\begin{equation}
        S(\tau,\phi)=
        -\Tr[\rho_{\rm at}\log_2\rho_{\rm at}]
        \le 1\ \text{ebit}.
\end{equation}
For numerical evaluations, the entropy is obtained by
diagonalizing the $2\times2$ atomic reduced state; the corresponding eigenvalue
formula is given in Appendix\ref{app:reduced}. For the
parameter range shown in Fig.~\ref{fig:qfit}(b), the entropy approaches the
one-ebit bound. The Fock cutoff and Poisson-tail tolerance are specified in
Appendix~\ref{app:numerics}.

%%%%%%%%%%%%%%%%%%%%%%%%%%%%%%%%%%%%%%%%%%%%%%%%%%%%%%%%%%%%%%%%%%%%%%%%%%
%%%%%%%%%%%%%%%%%%%%%%%%%%%%%%%%%%%%%%%%%%%%%%%%%%%%%%%%%%%%%%%%%%%%%
\section{Joint and reduced-state quantum Fisher information and readout}
\label{sec:qfi_readout}
%%%%%%%%%%%%%%%%%%%%%%%%%%%%%%%%%%%%%%%%%%%%%%%%%%%%%%%%%%%%%%%%%%%%%
This section derives the joint atom-cavity quantum Fisher information, then
uses the reduced states
$\rho_{\rm at}(\phi)=\Tr_{\rm cav}\rho_{ac}(\phi)$ and
$\rho_{\rm cav}(\phi)=\Tr_{\rm at}\rho_{ac}(\phi)$ to define the
atom-only and cavity-only reduced-state QFIs
$F_{\phi\phi}^{\rm at}\equiv F_Q[\rho_{\rm at}(\phi)]$ and
$F_{\phi\phi}^{\rm cav}\equiv F_Q[\rho_{\rm cav}(\phi)]$, and finally
compares practical readouts through their classical Fisher information (CFI).

\subsection{Joint and reduced-state QFI}

{
In the displaced-frame atom-cavity sensing protocol, the
force-canceling mechanical displacement leaves the motional
factor fixed as $\ket{0_m}$, so the estimated state is
$\ket{\Psi_{\rm disp}(g)}=\ket{0_m}\otimes\ket{\psi_{ac}(g)}$. Because
$\ket{0_m}$ is independent of $g$ in this model, the atom-cavity QFI is the
pure-state QFI of $\ket{\psi(t;g)}$ \cite{BraunsteinCaves1994,Paris2009}:
\begin{equation}
        F_{gg}^{ac}=4\left(
        \braket{\partial_g\psi}{\partial_g\psi}
        -|\braket{\psi}{\partial_g\psi}|^2
        \right).
\end{equation}
}
For the full lab-frame global QFI, the physical state contains
the $g$-dependent mechanical displacement,
$\ket{\Psi_{\rm lab}(g)}
=\widetilde D(\beta_g)\ket{0_m}\otimes\ket{\psi_{ac}(g)}$. Because this
global state factorizes into a mechanical displaced state and an atom-cavity
state, the corresponding product-state QFI separates into a
mechanical displacement term and the atom-cavity term $F_{gg}^{ac}$. The
mechanical contribution is derived in Appendix\ref{app:qfi},
Eq.\eqref{eq:app_lab_qfi_derivation}, and the lab-frame product-state QFI is
\begin{equation}
        F_{gg}^{\rm lab}=\frac{2m}{\hbar\omega_m^3}+F_{gg}^{ac}.
\end{equation}

The parameter enters only through $\Omega_n(g)=\Gc(g)\sqrt n$. The derivative
calculation, the norm of $\ket{\partial_g\psi}$, and the cancellation of
$\braket{\psi}{\partial_g\psi}$ are given in Appendix\ref{app:qfi},
Eq.\eqref{eq:app_ac_qfi_derivation}. The main text therefore records the final
QFI with respect to the physical acceleration $g$ and the
corresponding dimensionless QFI with respect to $\phi=kg/\omega_m^2$. Since
$\partial_g\Omega_n=\sqrt n\,\partial_g\Gc(g)$ and
$\sum_n n|c_n|^2=\bar n=|\alpha|^2$ for a coherent state,
\begin{equation}
        {
{
        F_{gg}^{ac}(t)=4|\alpha|^2t^2[\partial_g\Gc(g)]^2.
}
        }
\end{equation}
Using Eq.\eqref{eq:dGc},
\begin{equation}
{
        F_{gg}^{ac}(t)=
        4|\alpha|^2t^2
        \left(\Gmax\frac{k}{\omega_m^2}\right)^2
        \sin^2\!\left(kx_0-\frac{kg}{\omega_m^2}\right).
}
\end{equation}
Equivalently,
\begin{equation}
        {
        \Fphi(\tau,\phi)=4\bar n\tau^2\sin^2(\theta_0-\phi).
        }
\label{eq:Fphi}
\end{equation}
{
Since $\phi=kg/\omega_m^2$, the physical-acceleration and dimensionless QFIs are
related by the parameter-transformation rule
$F_{gg}^{ac}=\Fphi(k/\omega_m^2)^2$. At a node working point,
$\theta_0-\phi=\pi/2$, the responsivity is
$\left|\partial_g\Gc\right|_{\rm node}=\Gmax k/\omega_m^2$, so the node QFI is
\begin{equation}
        F_{gg,\mathrm{node}}^{ac}
        =
        4\bar n t^2
        \left(
        \Gmax\frac{k}{\omega_m^2}
        \right)^2.
\end{equation}
This node expression isolates the ideal scaling with $\bar n$, $t$, $\Gmax$,
$k$, and $\omega_m^{-2}$.
}
Equation\eqref{eq:Fphi} gives the displaced-frame atom-cavity
QFI that sets the precision bound
$\Delta\phi\ge1/\sqrt{\Fphi}$.
It is quadratic in interrogation time because the parameter controls a coherent
frequency. It is linear in $\bar n$, so a coherent probe gives
standard-quantum-limit scaling in photon number \cite{GiovannettiLloydMaccone2011}.  Finally, it is governed by the
square of the standing-wave slope: the sensitivity is maximal at a node and
zero at an antinode. For example, at $\bar n=9$ and $\tau=2\pi$ at the node, the
single-shot bound is $\Delta\phi\ge1/\sqrt{\Fphi}=1/(12\pi)=0.0265$.

\begin{figure}[t]
\centering
\includegraphics[width=\textwidth]{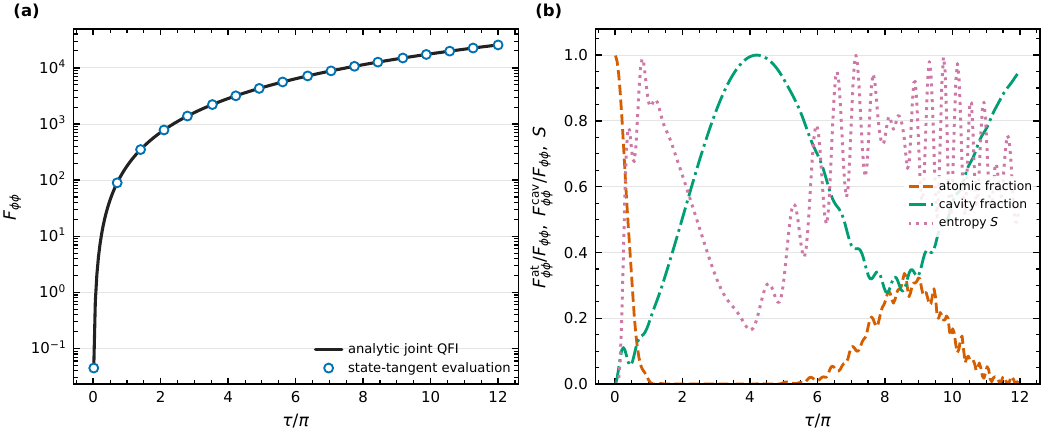}
\caption{\textbf{QFI versus time.}
(a) The joint atom-cavity QFI follows the exact parabola
$\Fphi=4\bar n\tau^2\sin^2(\theta_0-\phi)$; a numerical state-tangent evaluation reproduces the
closed form.
(b) Local atom-only and cavity-only QFI fractions,
$F_{\phi\phi}^{\rm at}/\Fphi$ and $F_{\phi\phi}^{\rm cav}/\Fphi$, together with
entanglement entropy $S$. The gravitational information is redistributed between
local subsystems while remaining bounded by the joint
atom-cavity QFI. The working point is
$\theta_0=\pi/2$, $\phi=\pi/4$, and $\bar n=9$.}
\label{fig:qfit}
\end{figure}
{
Panel (b) of Fig.\ref{fig:qfit} separates the joint precision bound from local
accessibility, quantified here by the atom-only and cavity-only
QFI fractions $F_{\phi\phi}^{\rm at}/\Fphi$ and
$F_{\phi\phi}^{\rm cav}/\Fphi$. The smooth curve in panel (a) gives the joint
atom-cavity QFI, which grows as $\tau^2$ even while local observables undergo
collapse and revival. Panel (b) shows the redistribution of  reduced-state QFI under
partial trace: at early times the atomic reduced state
$\rho_{\rm at}$ contains most of the  reduced-state QFI, whereas at later times
the cavity reduced state $\rho_{\rm cav}$ carries an
increasing fraction through its photon-number structure. The entropy curve
demonstrates that  reduced-state QFI need not track atom-cavity entanglement
monotonically: entanglement redistributes accessibility between subsystems but
does not change the joint bound
$\Fphi=4\bar n\tau^2\sin^2(\theta_0-\phi)$.
}The same distinction between the joint QFI bound and the
locally accessible information in reduced states can be visualized directly in
the $(\tau,\phi)$ plane. These plots are numerical evaluations of the
exact closed-form state and reduced density matrices.

\begin{figure}[t]
\centering
\includegraphics[width=\textwidth]{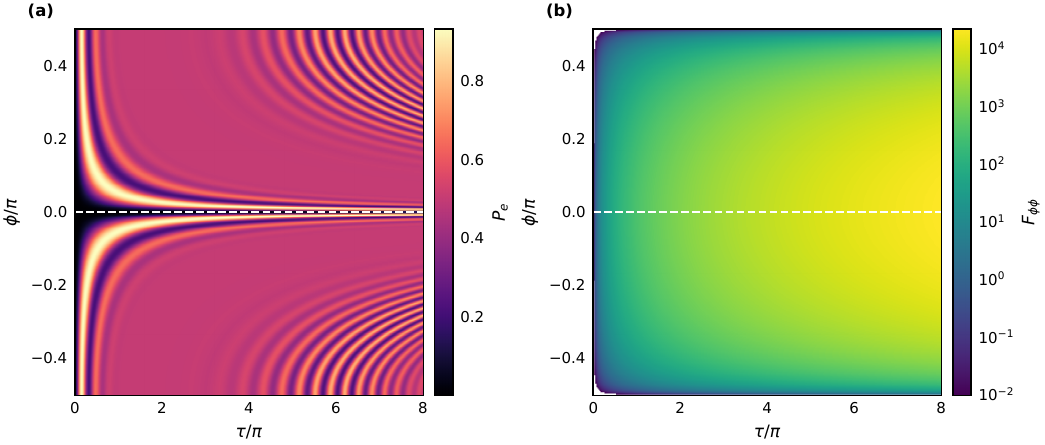}
\caption{\textbf{Signal and joint atom-cavity sensitivity maps.}
(a) Atomic excited-state population $P_e(\tau,\phi)$ computed from the exact JC
state for $\theta_0=\pi/2$ and $\bar n=9$. The node
$\phi=0$ is dark in population because the carrier coupling vanishes there,
whereas away from the node the fringes accelerate as
$\Omega_n=\Gmax\cos(\theta_0-\phi)\sqrt n$.
(b) Joint displaced-frame atom-cavity QFI
$F_{\phi\phi}=4\bar n\tau^2\sin^2(\theta_0-\phi)$, shown on a logarithmic color
scale for the same parameters. The QFI is maximal on the node ridge because it
is proportional to the square of the standing-wave slope, even though the
population signal in panel (a) vanishes at that point. Values
below $F_{\phi\phi}=10^{-2}$ are masked on the logarithmic display.}
\label{fig:landscape}
\end{figure}

{
Figure\ref{fig:landscape} compares the atomic population signal with the joint
displaced-frame atom-cavity QFI $\Fphi(\tau,\phi)$. Panel
(a) shows the excited-state population $P_e(\tau,\phi)$. At the node, $\phi=0$
for $\theta_0=\pi/2$, the carrier coupling vanishes and the population signal is
dark. Away from the node, the oscillations accelerate because, at node bias,
$u(\phi)=\sin\phi$ and therefore $\Omega_n=\Gmax\sin\phi\sqrt n$. Panel (b)
shows the atom-cavity QFI,
$\Fphi=4\bar n\tau^2\sin^2(\theta_0-\phi)$, which is maximal at the node.
Thus the population signal and the slope-controlled QFI are complementary: the
population signal is small where the standing-wave slope is largest.
}

\begin{figure}[t]
\centering
\includegraphics[width=\textwidth]{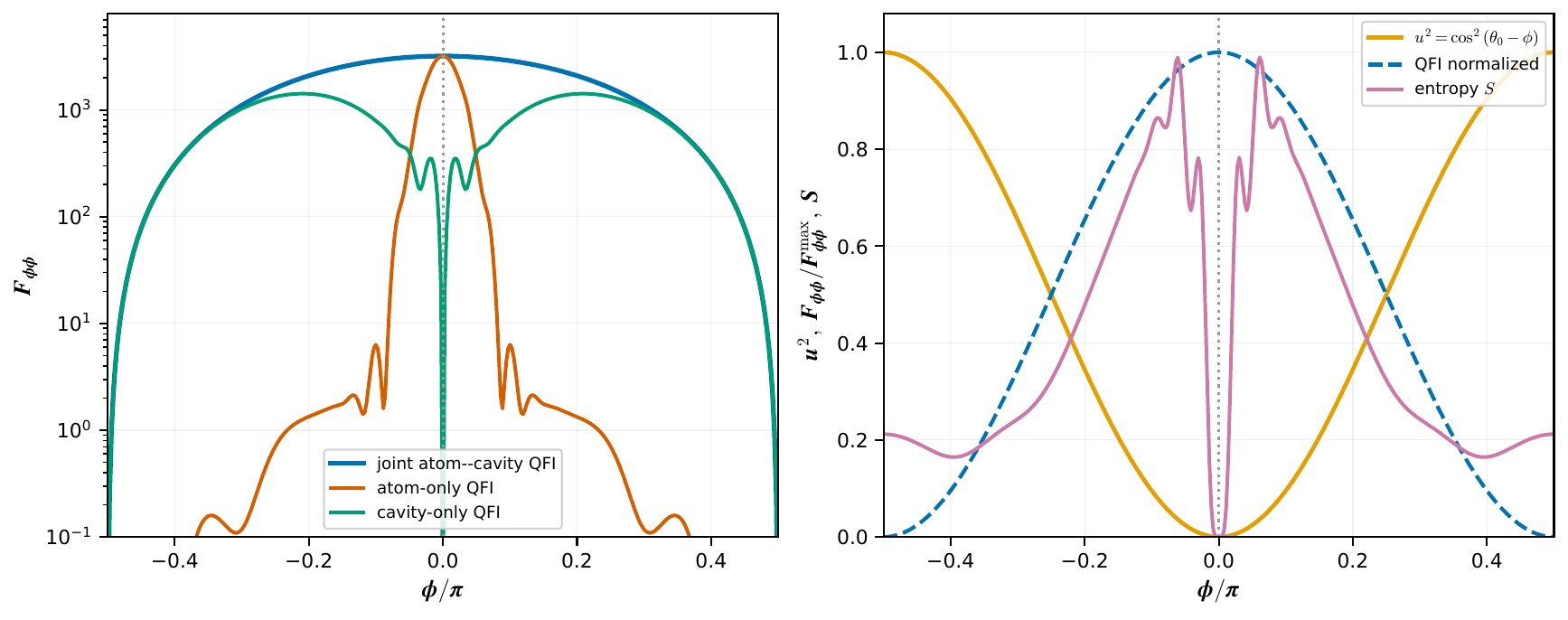}
\caption{\textbf{Atom--cavity local-QFI hierarchy versus
local signal phase.}
(a) Joint atom--cavity QFI and local atom-only and cavity-only
QFIs at $\tau=3\pi$, $\theta_0=\pi/2$, and $\bar n=9$.
At the node, the atom-only QFI saturates the joint atom--cavity bound while the
cavity-only QFI vanishes. Away from the node, the cavity-only
QFI becomes the dominant local contribution.
(b) Node complementarity for the same parameters: the squared carrier coupling
$u^2=\cos^2(\theta_0-\phi)$, the normalized joint atom--cavity QFI, and the
atom--cavity entanglement entropy $S$ in ebits. The QFI is largest where the
carrier coupling and entanglement are smallest, showing that the node sensitivity
comes from the first-order slope of the standing wave rather than from strong
atom--cavity entanglement.}
\label{fig:qfig}
\end{figure}

The same complementarity appears in the local signal-phase scan of
Fig.\ref{fig:qfig}. For the representative point used in the figure,
$\tau=3\pi$ and $\bar n=9$, the node value is
$\Fphi=4\bar n\tau^2=4(9)(3\pi)^2
=324\pi^2\simeq3197.75$, with
$F_{\phi\phi}^{\rm at}=\Fphi\simeq3197.75$ and
$F_{\phi\phi}^{\rm cav}=0$. The atomic reduced state saturates the joint
atom--cavity bound because, to first order in $\phi$ at the node, the state
{factorizes as
$\ket{\psi(\tau;\phi)}
\simeq
\left(\ket g-i\tau\alpha\phi\ket e\right)\otimes\ket\alpha$}. Off node, the cavity reduced state carries a finite fraction of the QFI\@. At the
representative off-node point $\phi=0.2\pi$ and $\tau=3\pi$, the corresponding
local-QFI values are
$F_{\phi\phi}^{\rm at}\simeq1.3$,
$F_{\phi\phi}^{\rm cav}\simeq1417$, and
$\Fphi\simeq2093$. Thus the locally most informative
subsystem changes with operating point.

The data-processing inequality requires  reduced-state QFIs to obey \cite{PetzGhinea2011,Ferrie2014}
\begin{equation}
        F_{\phi\phi}^{\rm at}\le \Fphi,
        \qquad
        F_{\phi\phi}^{\rm cav}\le \Fphi.
\end{equation}
The numerical implementation of the reduced-state QFIs and
readout CFIs, including the cutoff, tangent, spectral-support, grid, and
tolerance conventions, is specified in Appendix~\ref{app:numerics}.
\subsection{Classical Fisher information of practical readouts}
\begin{figure}[h]
\centering
\includegraphics[width=\textwidth]{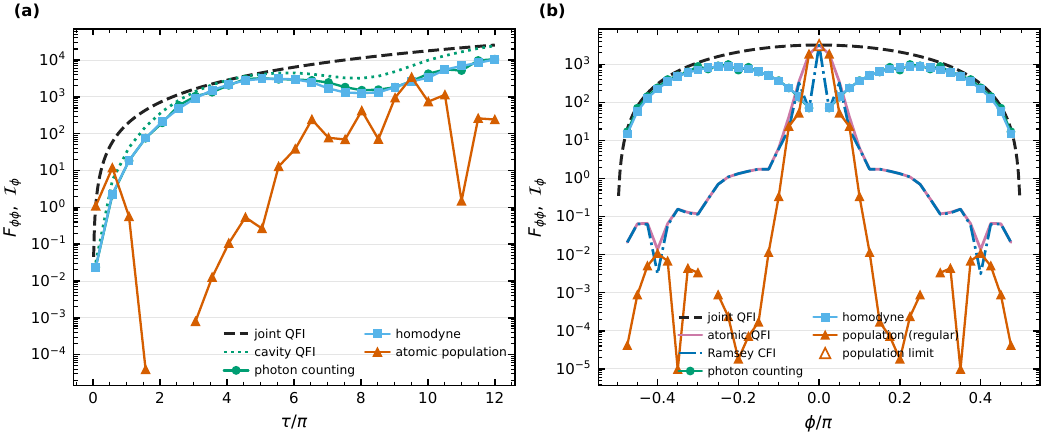}
\caption{\textbf{Classical Fisher information of practical readouts.}
(a) Off-node readout comparison at
$\phi=\pi/4$, $\theta_0=\pi/2$, and $\bar n=9$. The dashed
curve is the joint atom--cavity QFI, the dotted curve is the cavity-only QFI,
and the connected curves show the CFIs of photon counting, phase-optimized homodyne detection,
and atomic population readout. The ratios of the cavity-readout CFIs to the cavity-only QFI
vary appreciably with interrogation time.
(b) Local signal-phase scan at $\tau=3\pi$ for the same
$\theta_0$ and $\bar n$. The atomic QFI/SLD bound and the phase-referenced Ramsey CFI for
$\chi=\arg\alpha-\pi/2$ are shown separately. At the node, cavity photon counting and homodyne are
first-order insensitive to $\phi$. Atomic population readout is
nonregular and sign-insensitive at the exact node: its one-sided limiting FI is
finite, but $P_e(\phi)=P_e(-\phi)$ and the regular score formula is singular at
$P_e=0$. The open population marker denotes this one-sided nonregular limit,
whereas the connected regular population-CFI curve excludes the singular node. The phase-referenced Ramsey measurement of a transverse atomic
pseudospin component, which is sensitive to $\rho_{eg}$, is the regular
sign-sensitive readout that saturates the joint atom--cavity bound at the node;
off-node the cavity readouts dominate.}
\label{fig:read}
\end{figure}

The QFI bounds the precision over all possible measurements. A fixed measurement
defines the CFI
\begin{equation}
        \Icl_\phi=
        \sum_x \frac{[\partial_\phi p(x|\phi)]^2}{p(x|\phi)},
        \qquad
        \Icl_\phi\le F_{\phi\phi},
\label{eq:CFI_def}
\end{equation}
for the probabilities $p(x|\phi)$ of its outcomes. The readout models used in
Fig.\ref{fig:read} are organized by local subsystem and measured observable.
Photon counting uses the cavity probabilities
$p(n|\phi)=\bra n\rho_{\rm cav}(\phi)\ket n$. Homodyne detection at
local-oscillator phase $\vartheta$ uses
$p_\vartheta(x|\phi)=\bra{x_\vartheta}\rho_{\rm cav}(\phi)\ket{x_\vartheta}$,
with the CFI optimized over $\vartheta$; the Fock-basis expression used in the
numerics is given in Appendix\ref{app:cfi},
Eq.\eqref{eq:homodyne_fock_probability}. Atomic population readout uses
\begin{equation}
        \Icl_{\sigma_z}=
        \frac{[\partial_\phi \rho_{ee}(\phi)]^2}
        {\rho_{ee}(\phi)[1-\rho_{ee}(\phi)]}.
\end{equation}
Finally, a phase-referenced Ramsey measurement with analysis phase $\chi=\arg\alpha-\pi/2$ measures
a transverse atomic pseudospin component, equivalently a
rotated-basis observable sensitive to the coherence $\rho_{eg}$.

Figure\ref{fig:read} compares the CFI of these readouts with the relevant
QFI bounds. For the plotted off-node working point, the cavity reduced state
carries most of the reduced-state QFI, and cavity photon counting and phase-optimized homodyne detection recover time-dependent fractions of the cavity-only QFI bound, as shown explicitly in Fig.~\ref{fig:read}(a). Atomic
population readout without Ramsey analysis recovers a smaller fraction of the
corresponding local-QFI bound at the same off-node point.
At the node, the atom-only QFI becomes the relevant local bound.
The cavity state is first-order insensitive to $\phi$, so photon counting
and homodyne detection have zero first-order CFI. Atomic population readout
is nonregular at the exact node. For $\theta_0=\pi/2$,
$u(\phi)=\sin\phi$, and the small-$\phi$ expansion gives
\begin{subequations}
\label{eq:node_population_expansion}
\begin{align}
        u(\phi)
        &=
        \phi+O(\phi^3),
\label{eq:node_u_expansion}
        \\
        P_e(\tau;\phi)
        &=
        \sum_{n=1}^{\infty}
        |c_n|^2
        \sin^2\!\bigl(\sqrt n\,u(\phi)\tau\bigr)
\nonumber
        \\
        &=
        \bar n\tau^2\phi^2+O(\phi^4),
\label{eq:node_population_quadratic}
        \\
        \partial_\phi P_e(\tau;\phi)
        &=
        2\bar n\tau^2\phi+O(\phi^3).
\label{eq:node_population_derivative}
\end{align}
\end{subequations}
For $\phi\ne0$, the corresponding population CFI is
\begin{equation}
\begin{aligned}
        \Icl_{\sigma_z}(\tau;\phi)
        &=
        \frac{
        [\partial_\phi P_e(\tau;\phi)]^2
        }{
        P_e(\tau;\phi)[1-P_e(\tau;\phi)]
        }
        \\
        &=
        4\bar n\tau^2+O(\phi^2),
\end{aligned}
\label{eq:node_population_CFI}
\end{equation}
and therefore
\begin{equation}
        \lim_{\phi\to0}
        \Icl_{\sigma_z}(\tau;\phi)
        =
        4\bar n\tau^2 .
\label{eq:node_population_CFI_limit}
\end{equation}
The probability therefore vanishes quadratically at the node:
the one-sided limiting binary FI is finite, but the regular score formula is
singular at the exact point $P_e=0$. Population readout is also
sign-insensitive because $P_e(\phi)=P_e(-\phi)$. The regular,
sign-sensitive node readout is therefore a Ramsey measurement of a transverse
atomic pseudospin component: it is sensitive to the coherence $\rho_{eg}$,
whereas population readout measures the even quantity $P_e$.

The readout choice follows from the ordering of the atom-only
and cavity-only QFIs, $F_{\phi\phi}^{\rm at}$ and
$F_{\phi\phi}^{\rm cav}$. At the node, and for
sign-sensitive estimation in its immediate neighborhood, the appropriate local
measurement is a Ramsey measurement of the transverse atomic pseudospin
component selected by the phase-referenced analysis phase $\chi=\arg\alpha-\pi/2$. Away from the node, once
$F_{\phi\phi}^{\rm cav}>F_{\phi\phi}^{\rm at}$, cavity
photon counting or optimized homodyne becomes the preferred local readout.
Atomic population readout without Ramsey analysis is not a
regular sign-sensitive node readout: at the exact node it is nonregular and even
in $\phi$, and the population is only quadratically sensitive to $\phi$.
\section{Open-system dynamics and optimal interrogation}
\label{sec:open}
The closed-system QFI increases as $\tau^2$. In an experiment, cavity
loss, atomic decay, and dephasing limit the interrogation time
$t$, or equivalently $\tau=\Gmax t$. At the representative off-node working point considered below, cavity loss produces a finite QFI-maximizing time
$\tau^\ast$ \cite{HuelgaEtAl1997,EscherMatosFilhoDavidovich2011,
DemkowiczDobrzanskiKolodynskiGuta2012}. A minimal Markovian description
with cavity loss, atomic decay, and dephasing is the Lindblad
equation \cite{Lindblad1976,GoriniKossakowskiSudarshan1976}
\begin{equation}
{
\frac{d\rho}{dt}=
-\frac{i}{\hbar}[\hat H_{\rm eff}(g),\rho]
+\kappa_{\rm ph}\Dop[a]\rho
+\gamma_{\rm ph}\Dop[\sigma_-]\rho
+\frac{\gamma_{\phi,{\rm ph}}}{2}\Dop[\sigma_z]\rho,
}
\label{eq:lindblad}
\end{equation}
where
$\Dop[L]\rho=L\rho L^\dagger-\frac12\{L^\dagger L,\rho\}$. In dimensionless
time $\tau=\Gmax t$, with
$\tilde\kappa=\kappa_{\rm ph}/\Gmax$,
$\tilde\gamma=\gamma_{\rm ph}/\Gmax$, and
$\tilde\gamma_\phi=\gamma_{\phi,{\rm ph}}/\Gmax$, Eq.\eqref{eq:lindblad}
becomes
\begin{equation}
{
\frac{d\rho}{d\tau}
=
-i\left[\frac{\hat H_{\rm eff}(g)}{\hbar\Gmax},\rho\right]
+\tilde\kappa\Dop[a]\rho
+\tilde\gamma\Dop[\sigma_-]\rho
+\frac{\tilde\gamma_\phi}{2}\Dop[\sigma_z]\rho.
}
\label{eq:lindblad_dimless}
\end{equation}
The effective open-system reduction requires
dissipation not to wash out the carrier--sideband separation
set by $\omega_m$. In addition to Eq.\eqref{eq:rsb}, the
rates must satisfy
$\omega_m\gg\kappa_{\rm ph},\gamma_{\rm ph},\gamma_{\phi,{\rm ph}}$, or
equivalently
$\omega_m/\Gmax\gg\tilde\kappa,\tilde\gamma,\tilde\gamma_\phi$.

\begin{figure}[t]
\centering
\includegraphics[width=\textwidth]{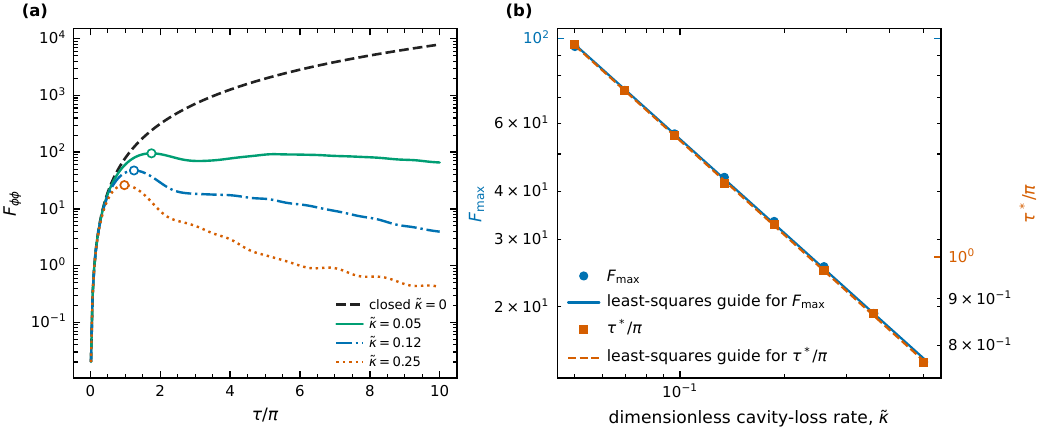}
\caption{\textbf{Representative open-system QFI.}
(a) Mixed-state joint atom-cavity QFI versus dimensionless time
for the Lindblad model in Eq.\eqref{eq:lindblad_dimless}. The dashed curve is
the closed-system result, and the solid curves show representative cavity-loss
rates $\tilde\kappa=0.05$, $0.12$, and $0.25$, with
$\theta_0=\pi/2$, $\phi=\pi/4$, $\bar n=4$, and
$\tilde\gamma=\tilde\gamma_\phi=0$. Markers indicate the loss-limited optimal
interrogation times.
(b) Peak QFI and optimal time over eight logarithmically
spaced values of $\tilde\kappa\in[0.05,0.5]$. The fitted lines shown in the
panel are descriptive finite-range summaries of the displayed numerical data
and are not interpreted as asymptotic or universal scaling laws.}
\label{fig:open}
\end{figure}

For a mixed state $\rho(\phi)=\sum_j\lambda_j\ket j\bra j$, the QFI is computed from the standard spectral expression \cite{LiuYuanLuWang2020,Safranek2018}
\begin{equation}
        F_{\phi\phi}=
        2\sum_{j,k:\lambda_j+\lambda_k>0}
        \frac{|\bra j\partial_\phi\rho\ket k|^2}{\lambda_j+\lambda_k}.
\label{eq:mixed_QFI}
\end{equation}
The open-system mixed-state QFI curves and the extracted
values of $F_{\max}$ and $\tau^\ast$ are shown in Fig.\ref{fig:open}.
At the representative off-node working point $\phi=\pi/4$,
cavity loss converts the closed-system $\tau^2$ growth into a finite maximum
$F_{\max}$ attained at an optimal interrogation time $\tau^\ast$.
Figure~\ref{fig:open}(b) reports the extracted values over
$\tilde\kappa\in[0.05,0.5]$. Least-squares guide lines summarize their
finite-range variation and are not interpreted as asymptotic scaling laws.
The finite optimum results from the combined action of cavity loss and the
multi-frequency Jaynes--Cummings dynamics. The values of $F_{\max}$ and
$\tau^\ast$ are obtained by tangent-Liouvillian propagation and local
shape-preserving peak refinement; the numerical details are given in
Appendix~\ref{app:numerics}.

%%%%%%%%%%%%%%%%%%%%%%%%%%%%%%%%%%%%%%%%%%%%%%%%%%%%%%%%%%%%%%%
%%%%%%%%%%%%%%%%%%%%%%%%%%%%%%%%%%%%%%%%%%%%%%%%%%%%%%%%%%%%%%%%%%%%%
\section{Discussion and outlook}
%%%%%%%%%%%%%%%%%%%%%%%%%%%%%%%%%%%%%%%%%%%%%%%%%%%%%%%%%%%%%%%%%%%%%

{
{
The central result is the following frequency-domain
transduction chain: the static
force shifts the trap equilibrium, the standing wave converts that shift into
the retained leading Lamb Dicke carrier coupling $G_c(g)$, and the JC ladder
converts $G_c(g)$ into exchange frequencies $\Omega_n(g)=G_c(g)\sqrt n$. This
is distinct from free-fall phase accumulation, direct displacement readout, and
dispersive photon-number shifts. The exact solution shows that the resulting
distinguishability is governed by the standing-wave slope rather than by the
magnitude of the excited-state population signal.
}
}

{
The node working point illustrates this distinction. At the node the carrier
coupling and excited-state population are small, but the local derivative
$\partial_gG_c$ is maximal. The appropriate subsystem-local
measurement is therefore not determined by the largest population contrast: near
the node it is an atomic Ramsey measurement, whereas away from the node it is a
cavity measurement. Near the node, the local information
about $\phi$ resides in the atomic coherence and is accessed by a Ramsey
measurement of a transverse pseudospin component; away from the node, the
cavity state becomes the relevant local object for photon counting or optimized
homodyne detection.
}

In the present single-atom coherent-probe protocol,
atom-cavity entanglement redistributes the  reduced-state QFI between
the atom-only and cavity-only reduced states but does not change
the joint atom-cavity scaling $\Fphi\propto\bar n$. At the node, maximal slope
sensitivity can occur with vanishing carrier coupling and vanishing atom-cavity
entanglement, so the sensitivity originates from a small coherent atomic rotation
driven by the coherent field amplitude. The protocol therefore remains
shot-noise limited in coherent-probe photon number. This conclusion applies to
the single-atom coherent-probe protocol; it does not preclude
scaling beyond $\Fphi\propto\bar n$ from entangled input
states or collective extensions. {
The validity regime of the carrier-only approximation is set
by the omitted phonon-assisted sideband $G_1(b+b^\dagger)K$, which is generated
by the same standing-wave slope that provides responsivity.
The Lamb Dicke support condition and the sideband diagnostics
$\epsilon_{\rm LD}$, $\epsilon_1$, and $\epsilon_2$
in Sec.\ref{sec:model} are therefore
validity conditions for the controlled effective model, not
resources that increase the QFI or readout signal.
A physical implementation would also require apparatus-specific values of the
wavelength, trap frequency, coupling strength, photon number, and loss rates.
The dimensionless examples in the figures are illustrative of the effective
theory and do not constitute an apparatus-level comparison to state-of-the-art
gravimeters.
}

{
Changing the coherent-probe photon-number scaling
$\Fphi\propto\bar n$ would require ingredients outside the
single-particle coherent-probe model solved here, such as nonclassical input
fields, collective atomic squeezing, sideband-based force sensing, or a
different mechanical transducer. Such extensions modify either the probe
resource or the force-to-carrier transduction mechanism
rather than only the final readout.
}
%%%%%%%%%%%%%%%%%%%%%%%%%%%%%%%%%%%%%%%%%%%%%%%%%%%%%%%%%%%%%%%%%
\section{Conclusion}
%%%%%%%%%%%%%%%%%%%%%%%%%%%%%%%%%%%%%%%%%%%%%%%%%%%%%%%%%%%%%%%%%%%%%

A constant gravitational acceleration shifts the equilibrium of the trapped
center-of-mass oscillator. This shift changes the local standing-wave phase at the atom’s equilibrium position. The resulting phase change modifies the carrier coupling. The acceleration-dependent carrier coupling therefore sets the Jaynes--Cummings exchange frequencies. The exact closed-system solution gives the quantum Fisher information of the joint atom--cavity state. It also shows how much of this information is retained in the atomic and cavity reduced states.  The QFI is proportional to the square of
the standing-wave slope and, for a coherent probe, grows quadratically with
interrogation time and linearly with mean photon number. Near a standing-wave
node, a phase-referenced atomic Ramsey measurement provides a regular,
sign-sensitive readout. At off-node operating points where the cavity reduced
state carries more quantum Fisher information than the atomic state, photon
counting and optimized homodyne detection provide explicit cavity readouts.
At the off-node working point considered in the open-system analysis, cavity
loss changes the quadratic quantum Fisher information growth into a maximum
reached at a finite interrogation time. The carrier approximation requires the occupied motional states to remain within the Lamb--Dicke regime and the phonon-assisted sideband corrections to remain small. After tracing out the motional mode, the unexpanded closed-system model reproduces the atom--cavity quantum Fisher information predicted by the carrier model throughout the simulated interrogation interval. The local standing-wave slope determines how strongly the carrier coupling responds to gravity. The same slope also controls the leading phonon-assisted sidebands and therefore enters the validity conditions of the carrier approximation.
\section*{Acknowledgement}
This publication was partially supported by the Qatar Research, Development and Innovation (QRDI) Council under the Academic Research Grant ARG01-0603-230468. The findings and views expressed herein are solely the responsibility of the authors.

\clearpage
\bibliographystyle{unsrtnat}
\bibliography{references}
%%%%%%%%%%%%%%%%%%%%%%%%%%%%%%%%%%%%%%%%%%%%%%%%%%%%%%%%%%%%%%%%%%%%%
\appendix
\renewcommand{\thesection}{\Alph{section}}
\renewcommand{\thesubsection}{\thesection.\arabic{subsection}}
\titleformat{\section}{\normalfont\scshape\centering}{Appendix\thesection:}{0.6em}{}
\titlespacing*{\section}{0pt}{1.4em}{0.6em}
\titleformat{\subsection}{\normalfont\bfseries}{\thesubsection}{0.7em}{}
\titlespacing*{\subsection}{0pt}{1.0em}{0.35em}
%%%%%%%%%%%%%%%%%%%%%%%%%%%%%%%%%%%%%%%%%%%%%%%%%%%%%%%%%%%%%%%%%%%%%
\section{Exact displacement of the forced oscillator}
\label{app:disp}
%%%%%%%%%%%%%%%%%%%%%%%%%%%%%%%%%%%%%%%%%%%%%%%%%%%%%%%%%%%%%%%%%%%%%

The purely mechanical part of Eq.\eqref{eq:Hlab} is
\begin{equation}
        \hat H_m(g)=
        \hbar\omega_m b^\dagger b
        +mgx_0
        +mgx_{\rm zpf}(b+b^\dagger).
\end{equation}
Define
\begin{equation}
        D(\beta_g)=\exp[\beta_g(b^\dagger-b)],
        \qquad
        \beta_g=\frac{mgx_{\rm zpf}}{\hbar\omega_m}.
\end{equation}
Let $A=\beta_g(b^\dagger-b)$. The BCH formula gives
\begin{equation}
        D^\dagger bD=e^{-A}be^{A}
        =b+[-A,b]+\frac{1}{2!}[-A,[-A,b]]+\cdots.
\end{equation}
Using $[b,b^\dagger]=1$, the two required adjoint actions are
\begin{equation}
\begin{aligned}
D^\dagger bD
&=b-\beta_g[b^\dagger-b,b]
  +\frac{1}{2!}[-A,[-A,b]]+\cdots
 =b+\beta_g,
\\
D^\dagger b^\dagger D
&=b^\dagger-\beta_g[b^\dagger-b,b^\dagger]
  +\frac{1}{2!}[-A,[-A,b^\dagger]]+\cdots
 =b^\dagger+\beta_g,
\end{aligned}
\end{equation}
where the nested commutators vanish because the first commutators are
$c$-numbers. The displacement that cancels the force in the main text is
$\widetilde D=D^\dagger$, so
\begin{equation}
        \widetilde D^\dagger b\widetilde D=b-\beta_g,
        \qquad
        \widetilde D^\dagger b^\dagger\widetilde D=b^\dagger-\beta_g.
\label{eq:app_disp_rules}
\end{equation}
Substitution into the mechanical Hamiltonian gives
\begin{equation}
\begin{aligned}
\widetilde D^\dagger\hat H_m\widetilde D
&=\hbar\omega_m(b^\dagger-\beta_g)(b-\beta_g)
+mgx_0
+mgx_{\rm zpf}[(b+b^\dagger)-2\beta_g]
\\
&=\hbar\omega_m b^\dagger b
+[-\hbar\omega_m\beta_g+mgx_{\rm zpf}](b+b^\dagger)
+\hbar\omega_m\beta_g^2+mgx_0-2mgx_{\rm zpf}\beta_g.
\end{aligned}
\end{equation}
The linear term vanishes because $\hbar\omega_m\beta_g=mgx_{\rm zpf}$. The
remaining scalar term is a global phase. The transformed position and the
corresponding equilibrium shift are
\begin{equation}
\begin{aligned}
        \widetilde D^\dagger\hat x\widetilde D
        &=
        x_0+x_{\rm zpf}(b+b^\dagger)-2x_{\rm zpf}\beta_g,
\\
        x_{\rm eq}(g)
        &=-2x_{\rm zpf}\beta_g
        =-2x_{\rm zpf}\frac{mgx_{\rm zpf}}{\hbar\omega_m}
        =-\frac{2mg}{\hbar\omega_m}\frac{\hbar}{2m\omega_m}
        =-\frac{g}{\omega_m^2}.
\end{aligned}
\label{eq:app_xeq_derivation}
\end{equation}

%%%%%%%%%%%%%%%%%%%%%%%%%%%%%%%%%%%%%%%%%%%%%%%%%%%%%%%%%%%%%%%%%%%%%
\section{Optical rotating-wave, Lamb Dicke, and sideband reductions}
\label{app:LD}
%%%%%%%%%%%%%%%%%%%%%%%%%%%%%%%%%%%%%%%%%%%%%%%%%%%%%%%%%%%%%%%%%%%%%

The optical rotating-wave approximation used in Eq.\eqref{eq:Hlab} follows by
expanding the pre-RWA single-mode dipole interaction and comparing the
interaction-picture rotation frequencies. With
$H_0=\hbar\omega_c a^\dagger a+(\hbar\omega_q/2)\sigma_z$, one has
\begin{equation}
{
\begin{aligned}
H_{\rm int}^{\rm Rabi}
&=\hbar G(\hat x)
\left(a\sigma_+ +a\sigma_-+a^\dagger\sigma_+ +a^\dagger\sigma_-\right),
\\
a\sigma_+(t)
&=a\sigma_+e^{i(\omega_q-\omega_c)t},
&
a^\dagger\sigma_-(t)
&=a^\dagger\sigma_-e^{-i(\omega_q-\omega_c)t},
\\
a\sigma_-(t)
&=a\sigma_-e^{-i(\omega_q+\omega_c)t},
&
a^\dagger\sigma_+(t)
&=a^\dagger\sigma_+e^{i(\omega_q+\omega_c)t}.
\end{aligned}
}
\label{eq:app_optical_rwa_terms}
\end{equation}
Thus, for $|\Delta|=|\omega_q-\omega_c|\ll\omega_q+\omega_c$ and light-matter
matrix elements small compared with the optical frequencies, the
counter-rotating terms average out and the retained interaction is
\begin{equation}
{
        H_{\rm int}^{\rm RWA}
        =
        \hbar G(\hat x)(a\sigma_+ + a^\dagger\sigma_-).
}
\end{equation}

In the displaced frame,
\begin{equation}
        G(\hat x)=\Gmax\cos[\theta_g+\eta(b+b^\dagger)],
        \qquad
        \theta_g=kx_0-\frac{kg}{\omega_m^2},
        \qquad
        \eta=kx_{\rm zpf}.
\end{equation}
The controlled small parameter is the operator fluctuation
$\eta X$ with $X=b+b^\dagger$. For a number state $\ket m$,
\begin{equation}
{
        \langle m|X^2|m\rangle
        =
        \langle m|b^2+b^{\dagger2}+bb^\dagger+b^\dagger b|m\rangle
        =
        2m+1.
}
\end{equation}
Thus a sufficient Lamb Dicke condition for support mainly on
$m\le m_{\rm eff}$ is
\begin{equation}
{
        \eta\sqrt{2m_{\rm eff}+1}\ll1.
}
\end{equation}
Using
\begin{equation}
        \cos(\theta_g+\eta X)
        =
        \cos\theta_g\cos(\eta X)-\sin\theta_g\sin(\eta X),
\end{equation}
with $X=b+b^\dagger$, and expanding
\begin{equation}
        \cos(\eta X)=1-\frac{\eta^2X^2}{2}+O(\eta^4),
        \qquad
        \sin(\eta X)=\eta X+O(\eta^3),
\end{equation}
one obtains
\begin{equation}
        G(\hat x)
        =
        \Gmax\cos\theta_g
        -\Gmax\eta\sin\theta_g(b+b^\dagger)
        +O(\eta^2).
\end{equation}
This gives Eq.\eqref{eq:LD_main}.

{
For completeness, the exact diagonal carrier matrix element can be written in
closed form. Using
\begin{equation}
        e^{i\eta(b+b^\dagger)}
        =
        e^{-\eta^2/2}e^{i\eta b^\dagger}e^{i\eta b},
\end{equation}
one obtains the standard Lamb Dicke/Debye--Waller result
\begin{equation}
        \bra m e^{i\eta(b+b^\dagger)}\ket m
        =
        e^{-\eta^2/2}L_m(\eta^2),
\end{equation}
and hence
\begin{equation}
        \bra m
        \Gmax\cos[\theta_g+\eta(b+b^\dagger)]
        \ket m
        =
        \Gmax e^{-\eta^2/2}L_m(\eta^2)\cos\theta_g.
\end{equation}
The effective Hamiltonian in the main text keeps the leading carrier
$\Gmax\cos\theta_g$ and the first-order sideband coefficient
$-\Gmax\eta\sin\theta_g$. The Debye--Waller carrier correction is therefore an
$O(\eta^2)$ term, consistent with the order at which other omitted
Lamb Dicke corrections enter \cite{LeibfriedBlattMonroeWineland2003,
WinelandMonroeItano1998}.
}

The sideband-elimination step follows from the interaction picture of
\begin{equation}
        H_{\rm free}
        =
        \hbar\omega_c a^\dagger a
        +\frac{\hbar\omega_q}{2}\sigma_z
        +\hbar\omega_m b^\dagger b.
\end{equation}
The required adjoint actions are
\begin{equation}
\begin{aligned}
        a(t)&=ae^{-i\omega_ct},
        &
        a^\dagger(t)&=a^\dagger e^{i\omega_ct},
        \\
        b(t)&=be^{-i\omega_mt},
        &
        b^\dagger(t)&=b^\dagger e^{i\omega_mt},
\end{aligned}
\end{equation}
and
\begin{equation}
        \sigma_+(t)=\sigma_+e^{i\omega_qt},
        \qquad
        \sigma_-(t)=\sigma_-e^{-i\omega_qt}.
\end{equation}
With detuning $\Delta=\omega_q-\omega_c$, the interaction-picture coupling is
\begin{equation}
\begin{aligned}
V_I(t)
&=
\hbar\Gc(g)
\left(a\sigma_+e^{i\Delta t}
+a^\dagger\sigma_-e^{-i\Delta t}\right)
\\
&\quad+
\hbar G_1(g)
\left(be^{-i\omega_mt}+b^\dagger e^{i\omega_mt}\right)
\left(a\sigma_+e^{i\Delta t}
+a^\dagger\sigma_-e^{-i\Delta t}\right).
\end{aligned}
\end{equation}
On resonance, $\Delta=0$, the sideband part is Eq.\eqref{eq:Vsb_main}. Its
first-order integrated contribution is
\begin{equation}
\begin{aligned}
\int_0^t ds\,V_{\rm sb}^{(I)}(s)
&=
\hbar G_1(g)
\left[
        b\frac{1-e^{-i\omega_mt}}{i\omega_m}
        +b^\dagger\frac{e^{i\omega_mt}-1}{i\omega_m}
\right]
\\
&\quad\times
(a\sigma_+ + a^\dagger\sigma_-).
\end{aligned}
\end{equation}
The prefactor scales as $|G_1|/\omega_m$, giving the resolved-sideband condition
in Eq.\eqref{eq:rsb}. The first neglected averaged correction
may be estimated by writing
\begin{equation}
{
        V_{\rm sb}^{(I)}(t)=H_{-1}e^{-i\omega_mt}+H_{+1}e^{i\omega_mt},
        \qquad
        H_{-1}=\hbar G_1 bK,
        \quad
        H_{+1}=\hbar G_1 b^\dagger K.
}
\label{eq:app_Hpm}
\end{equation}
Since $b,b^\dagger$ commute with $K$,
\begin{equation}
{
        [H_{-1},H_{+1}]
        =
        \hbar^2G_1^2[bK,b^\dagger K]
        =
        \hbar^2G_1^2K^2.
}
\end{equation}
Thus the averaged correction scale is
\begin{equation}
{
        H_{\rm sb}^{(2)}
        \sim
        \hbar\frac{G_1^2}{\omega_m}K^2,
        \qquad
        t\frac{G_1^2}{\omega_m}n_{\rm eff}\ll1
}
\label{eq:app_Hsb_scale}
\end{equation}
if this term is not retained explicitly. The separate
first-order micromotion/leakage bound contains the photon and phonon support
factor,
\begin{equation}
{
        \frac{|G_1|}{\omega_m}
        \sqrt{(n_{\rm eff}+1)(m_{\rm eff}+1)}
        \ll1.
}
\end{equation}
For the displaced motional ground-state protocol,
$m_{\rm eff}=0$.
%%%%%%%%%%%%%%%%%%%%%%%%%%%%%%%%%%%%%%%%%%%%%%%%%%%%%%%%%%%%%%%%%%%%%
%%%%%%%%%%%%%%%%%%%%%%%%%%%%%%%%%%%%%%%%%%%%%%%%%%%%%%%%%%%%%%%%%%%%%
%%%%%%%%%%%%%%%%%%%%%%%%%%%%%%%%%%%%%%%%%%%%%%%%%%%%%%%%%%%%%%%%%%%%%
\section{Exact Jaynes Cummings propagator}
\label{app:prop}
%%%%%%%%%%%%%%%%%%%%%%%%%%%%%%%%%%%%%%%%%%%%%%%%%%%%%%%%%%%%%%%%%%%%%

Let
\begin{equation}
        K=a\sigma_+ + a^\dagger\sigma_-,
        \qquad
        N=a^\dagger a+\sigma_+\sigma_-.
\end{equation}
Using
\begin{equation}
\begin{aligned}
        [a^\dagger a,a]&=-a,
        &
        [a^\dagger a,a^\dagger]&=a^\dagger,
        \\
        [\sigma_+\sigma_-,\sigma_+]&=\sigma_+,
        &
        [\sigma_+\sigma_-,\sigma_-]&=-\sigma_-,
\end{aligned}
\end{equation}
and the fact that atom and cavity operators commute,
\begin{equation}
\begin{aligned}
[N,K]
&=
[a^\dagger a,a\sigma_+]
+[a^\dagger a,a^\dagger\sigma_-]
+[\sigma_+\sigma_-,a\sigma_+]
+[\sigma_+\sigma_-,a^\dagger\sigma_-]
\\
&=
-a\sigma_+
+a^\dagger\sigma_-
+a\sigma_+
-a^\dagger\sigma_-
=0.
\end{aligned}
\label{eq:app_JC_conservation}
\end{equation}
Therefore the exchange preserves total excitation number. For $n\ge1$,
\begin{equation}
        K\ket{g,n}=\sqrt n\ket{e,n-1},
        \qquad
        K\ket{e,n-1}=\sqrt n\ket{g,n}.
\end{equation}
Thus, in the ordered basis $\{\ket{g,n},\ket{e,n-1}\}$,
\begin{equation}
        \left.K\right|_n=\sqrt n\,\tau_x,
\end{equation}
where $\tau_x$ denotes the Pauli $x$ matrix in this two-state manifold. Since
$\tau_x^2=\mathbb I$,
\begin{equation}
        \left.e^{-i\Gc tK}\right|_n
        =
        e^{-i\Omega_n t\tau_x}
        =
        \cos(\Omega_n t)\mathbb I
        -i\sin(\Omega_n t)\tau_x.
\end{equation}
Thus
\begin{equation}
\begin{aligned}
        U_I(t;g)\ket{g,n}
        &=
        \cos(\Omega_n t)\ket{g,n}
        -i\sin(\Omega_n t)\ket{e,n-1},
        \\
        U_I(t;g)\ket{e,n-1}
        &=
        \cos(\Omega_n t)\ket{e,n-1}
        -i\sin(\Omega_n t)\ket{g,n}.
\end{aligned}
\label{eq:app_UI_action}
\end{equation}
The state $\ket{g,0}$ is invariant because $K\ket{g,0}=0$. Combining this with
the free resonant evolution yields Eq.\eqref{eq:psi_full_phase}.

%%%%%%%%%%%%%%%%%%%%%%%%%%%%%%%%%%%%%%%%%%%%%%%%%%%%%%%%%%%%%%%%%%%%%
\section{Reduced states and entanglement entropy}
\label{app:reduced}
%%%%%%%%%%%%%%%%%%%%%%%%%%%%%%%%%%%%%%%%%%%%%%%%%%%%%%%%%%%%%%%%%%%%%

Starting from Eq.\eqref{eq:conditional_form}, the atom-cavity density operator
is
\begin{equation}
\begin{aligned}
\rho_{ac}
&=\ket g\bra g\otimes\ket{\phi_g}\bra{\phi_g}
+\ket e\bra e\otimes\ket{\phi_e}\bra{\phi_e}
\\
&\quad+
\ket g\bra e\otimes\ket{\phi_g}\bra{\phi_e}
+\ket e\bra g\otimes\ket{\phi_e}\bra{\phi_g}.
\end{aligned}
\end{equation}
Tracing over the atom removes the cross terms and gives
$\rho_{\rm cav}=\ket{\phi_g}\bra{\phi_g}+\ket{\phi_e}\bra{\phi_e}$. Inserting the
coefficients $A_n,E_n$ gives Eq.\eqref{eq:rho_cav_elements}. Tracing over the
cavity gives Eq.\eqref{eq:rho_at} with entries
Eqs.\eqref{eq:rhoee}--\eqref{eq:rhoeg}.

The eigenvalues of the atomic qubit follow from the determinant
condition. Using $\rho_{ee}+\rho_{gg}=1$,
\begin{equation}
{
\begin{aligned}
0
&=
\det(\rho_{\rm at}-\lambda I)
=
(\rho_{ee}-\lambda)(\rho_{gg}-\lambda)-|\rho_{eg}|^2,
\\
0
&=
\lambda^2-\lambda+\rho_{ee}\rho_{gg}-|\rho_{eg}|^2,
\\
\lambda_\pm
&=
\frac12\left[1\pm
\sqrt{(\rho_{ee}-\rho_{gg})^2+4|\rho_{eg}|^2}\right].
\end{aligned}
}
\end{equation}
The entanglement entropy of the pure atom-cavity state is
\begin{equation}
        S=-\lambda_+\log_2\lambda_+-\lambda_-\log_2\lambda_-.
\end{equation}
Since there are only two Schmidt coefficients, $S\le1$ ebit.

%%%%%%%%%%%%%%%%%%%%%%%%%%%%%%%%%%%%%%%%%%%%%%%%%%%%%%%%%%%%%%%%%%%%%
%%%%%%%%%%%%%%%%%%%%%%%%%%%%%%%%%%%%%%%%%%%%%%%%%%%%%%%%%%%%%%%%%%%%%
\section{Pure-state quantum Fisher information}
\label{app:qfi}
%%%%%%%%%%%%%%%%%%%%%%%%%%%%%%%%%%%%%%%%%%%%%%%%%%%%%%%%%%%%%%%%%%%%%

For the pure state in Eq.\eqref{eq:psi_full_phase}, the $g$-dependence enters
only through $\Omega_n(g)$. Differentiating the JC amplitudes gives the complete
atom-cavity pure-state QFI derivation:
\begin{subequations}
\label{eq:app_ac_qfi_derivation}
\begin{align}
\ket{\partial_g\psi}
&=
\sum_{n=1}^\infty c_ne^{-i\omega(n-\frac12)t}
\left[
-t(\partial_g\Omega_n)\sin(\Omega_nt)\ket{g,n}
-i t(\partial_g\Omega_n)\cos(\Omega_nt)\ket{e,n-1}
\right],
\\
\braket{\partial_g\psi}{\partial_g\psi}
&=
t^2\sum_{n=1}^\infty |c_n|^2(\partial_g\Omega_n)^2
[\sin^2(\Omega_nt)+\cos^2(\Omega_nt)]
=
t^2\sum_{n=1}^\infty |c_n|^2(\partial_g\Omega_n)^2,
\\
\braket{\psi}{\partial_g\psi}
&=
\sum_{n=1}^\infty |c_n|^2
\left[
-t(\partial_g\Omega_n)\cos(\Omega_nt)\sin(\Omega_nt)
+t(\partial_g\Omega_n)\sin(\Omega_nt)\cos(\Omega_nt)
\right]
=0,
\\
{F_{gg}^{ac}}
&{
=
4t^2\sum_{n=1}^\infty |c_n|^2(\partial_g\Omega_n)^2
=
4t^2(\partial_g\Gc)^2\sum_{n=1}^\infty n|c_n|^2
=
4|\alpha|^2t^2(\partial_g\Gc)^2.
}
\end{align}
\end{subequations}
Changing variables from $g$ to $\phi=kg/\omega_m^2$ gives
Eq.\eqref{eq:Fphi}.
For the full lab-frame global QFI one must also include the
mechanical displacement. Writing the mechanical factor as
$\ket{\chi_m(g)}=\widetilde D(\beta_g)\ket{0_m}$, the displacement generator
and its vacuum variance give
\begin{subequations}
\label{eq:app_lab_qfi_derivation}
{
\begin{align}
        \ket{\chi_m(g)}
        &=
        \widetilde D(\beta_g)\ket{0_m},
        \qquad
        \partial_g\ket{\chi_m}
        =
        -(\partial_g\beta_g)(b^\dagger-b)\ket{\chi_m},
\\
        P
        &=
        -i(b^\dagger-b),
        \qquad
        (\Delta P)^2=1,
\\
        F_{gg}^{m}
        &=
        4(\partial_g\beta_g)^2(\Delta P)^2
        =
        4\left(\frac{mx_{\rm zpf}}{\hbar\omega_m}\right)^2
        =
        \frac{2m}{\hbar\omega_m^3},
\\
        F_{gg}^{\rm lab}
        &=
        F_{gg}^{m}+F_{gg}^{ac}
        =
        \frac{2m}{\hbar\omega_m^3}
        +
        F_{gg}^{ac}.
\end{align}
}
\end{subequations}
The first term belongs to the global lab-frame product-state
model and is absent from an atom-cavity-only displaced-frame statistical
model.
%%%%%%%%%%%%%%%%%%%%%%%%%%%%%%%%%%%%%%%%%%%%%%%%%%%%%%%%%%%%%%%%%%%%%
\section{Reduced-state QFI and classical Fisher information}
\label{app:cfi}
%%%%%%%%%%%%%%%%%%%%%%%%%%%%%%%%%%%%%%%%%%%%%%%%%%%%%%%%%%%%%%%%%%%%%

For a mixed state $\rho(\phi)=\sum_j\lambda_j\ket j\bra j$, the QFI is
\begin{equation}
        F_{\phi\phi}[\rho]=
        2\sum_{j,k:\lambda_j+\lambda_k>0}
        \frac{|\bra j\partial_\phi\rho\ket k|^2}{\lambda_j+\lambda_k}.
\label{eq:local_mixed_QFI}
\end{equation}
Applying this expression to $\rho_{\rm at}$ and $\rho_{\rm cav}$ gives the local
atom-only and cavity-only QFIs plotted in Figs.\ref{fig:qfit} and\ref{fig:qfig}. These obey the data-processing inequalities
$F_{\phi\phi}^{\rm at},F_{\phi\phi}^{\rm cav}\le\Fphi$ because each reduced state
is obtained from the global pure state by a parameter-independent partial trace.

For a POVM $\{E_x\}$ with probabilities $p(x|\phi)=\Tr[\rho(\phi)E_x]$, the CFI
is Eq.\eqref{eq:CFI_def}. The measurement models used in the numerical readout
comparison are as follows.

Photon counting uses $E_n=\ket n\bra n$ on the cavity, so
$p_n=\rho_{nn}$. Homodyne detection at phase $\vartheta$ uses the quadrature
basis $\ket{x_\vartheta}$, with
\begin{equation}
        p_\vartheta(x)=\bra{x_\vartheta}\rho_{\rm cav}\ket{x_\vartheta}
        =\sum_{n,n'}\rho_{nn'}e^{i(n-n')\vartheta}
        \varphi_n(x)\varphi_{n'}(x),
\label{eq:homodyne_fock_probability}
\end{equation}
where $\varphi_n(x)$ are harmonic-oscillator quadrature wave functions, and the
CFI is maximized over $\vartheta$. Atomic population readout uses the projectors
$\ket e\bra e$ and $\ket g\bra g$, giving
\begin{equation}
        \Icl_{\sigma_z}=
        \frac{(\partial_\phi\rho_{ee})^2}{\rho_{ee}(1-\rho_{ee})}.
\end{equation}
For the atomic qubit written as
$\rho_{\rm at}=\frac12(\mathbb I+\mathbf r\cdot\boldsymbol\sigma)$, the
optimized qubit measurement gives
\begin{equation}
        F_{\phi\phi}^{\rm at}=|
        \partial_\phi\mathbf r|^2
        +\frac{(\mathbf r\cdot\partial_\phi\mathbf r)^2}{1-|\mathbf r|^2},
\end{equation}
for $|\mathbf r|<1$, with the pure-state limit obtained continuously. At the node, the first-order node expansion in
Sec.\ref{sec:qfi_readout} shows explicitly that the informative atomic
quadrature is transverse to $\sigma_z$.  The
population probability behaves as
\begin{equation}
{
        P_e=\bar n\tau^2\phi^2+O(\phi^4),
        \qquad
        \Icl_{\sigma_z}=\frac{(\partial_\phi P_e)^2}{P_e(1-P_e)}\to4\bar n\tau^2
        \quad(\phi\to0,\ \phi\ne0),
}
\end{equation}
but it is nonregular at $P_e=0$ and sign-insensitive because
$P_e(\phi)=P_e(-\phi)$. Ramsey readout is therefore the regular sign-sensitive
node measurement.

%%%%%%%%%%%%%%%%%%%%%%%%%%%%%%%%%%%%%%%%%%%%%%%%%%%%%%%%%%%%%%%%%%%%%
%%%%%%%%%%%%%%%%%%%%%%%%%%%%%%%%%%%%%%%%%%%%%%%%%%%%%%%%%%%%%%%%%%%%%
\section{Numerical implementation details}
\label{app:numerics}
%%%%%%%%%%%%%%%%%%%%%%%%%%%%%%%%%%%%%%%%%%%%%%%%%%%%%%%%%%%%%%%%%%%%%

The closed-system reduced-state QFIs in
Figs.~\ref{fig:qfit} and \ref{fig:qfig}, together with the QFIs and CFIs in
Fig.~\ref{fig:read}, were computed from normalized truncated coherent states
and analytically differentiated state tangents. The cavity Fock space retained
the states $n=0,\ldots,n_f-1$, with $n_f=45$ for the closed-system QFI
figures and $n_f=40$ for the readout figure. For $\bar n=9$, the
corresponding omitted Poisson tails are $1.12\times10^{-17}$ and
$2.86\times10^{-14}$.

The normalized truncated state and its analytic tangent were used to construct
$\rho$ and $\partial_\phi\rho$. Roundoff-level non-Hermitian components
were removed by symmetrization, and
$\Tr(\partial_\phi\rho)=0$ was preserved up to numerical roundoff. The
mixed-state QFI was evaluated from Eq.~\eqref{eq:local_mixed_QFI}, retaining
eigenvalue-pair terms for which
$\lambda_j+\lambda_k>
\max(10^{-14},10^{-12}\lambda_{\max})$, where $\lambda_{\max}$ is the
largest eigenvalue of the density matrix. The data-processing inequalities
$F_{\phi\phi}^{\rm at},F_{\phi\phi}^{\rm cav}\le\Fphi$ were satisfied at
all sampled time and phase points.

The homodyne CFI used $401$ quadrature points over $x\in[-10,10]$ and
$33$ uniformly spaced local-oscillator phases over
$\vartheta\in[0,\pi)$. The best coarse phase was refined using $21$ points
over a window extending by one coarse-grid spacing on either side. The readout
curves use $25$ time points over $\tau\in[0.25,12\pi]$ and $41$ phase
points over $\phi\in[-\pi/2,\pi/2]$; connecting lines guide the eye between
the calculated samples. In Figs.~\ref{fig:qfig} and \ref{fig:read}, values
below the logarithmic display floor $10^{-6}$ were masked rather than
replaced by finite values.

The open-system curves in Fig.~\ref{fig:open} use dimensionless time
$\tau=\Gmax t$ and the master equation in
Eq.~\eqref{eq:lindblad_dimless}. The cavity Fock space was truncated to
$n=0,\ldots,21$, corresponding to $n_f=22$ retained cavity states. For
$\bar n=4$, the omitted initial Poisson tail is
$3.46\times10^{-10}$. The density matrix and its parameter tangent were
propagated simultaneously using the tangent Liouvillian in
Appendix~\ref{app:liouvillian}; the plotted QFIs were therefore obtained
without finite-difference derivatives. The same spectral-support criterion
specified above was used for the open-system mixed-state QFI.

Figure~\ref{fig:open}(a) uses $220$ uniformly spaced points over
$\tau\in[0.05,10\pi]$. Figure~\ref{fig:open}(b) uses eight logarithmically
spaced values of $\tilde\kappa\in[0.05,0.5]$. For each loss rate, the peak
search uses $320$ uniformly spaced points over
$\tau\in[0.05,\max(8\pi,5/\tilde\kappa)]$.For each value of $\tilde\kappa$, the sampled QFI maximum is bracketed by
its immediately adjacent time points. A shape-preserving piecewise cubic
Hermite interpolant (PCHIP), constructed from the sampled maximum and the two
points on either side, is maximized within this bracket to estimate
$\tau^\ast$ and $F_{\max}$. At
$\tilde\kappa\simeq0.134$, the QFI obtained by direct
tangent-Liouvillian propagation at the estimated $\tau^\ast$ differs from
the interpolated $F_{\max}$ by less than $10^{-10}$ in relative terms.

The fitted lines in Fig.~\ref{fig:open}(b) are ordinary least-squares
regressions of $\log F_{\max}$ and $\log(\tau^\ast/\pi)$ against
$\log\tilde\kappa$, using the eight sampled loss rates over
$\tilde\kappa\in[0.05,0.5]$. They describe the numerical variation over this
finite interval; no asymptotic scaling is inferred.

The unexpanded-model convergence checks were performed at $\phi=0$ and
$\phi=\pi/4$ for $\tau/\pi=1,3,6,$ and $12$. Increasing the photon
dimension from $n_f=42$ to $46$ at fixed $n_m=9$ changes the
motion-traced atom--cavity QFI by a maximum relative amount of
$6.4\times10^{-13}$. Increasing the phonon dimension from $n_m=9$ to
$11$ at fixed $n_f=42$ changes the same QFI by a maximum relative amount
of $8.7\times10^{-13}$. At these eight $(\phi,\tau)$ points, centered
finite differences with steps from $5\times10^{-7}$ to
$2\times10^{-6}$ give global pure-state QFIs whose maximum relative
difference from the tangent-state result is $2.3\times10^{-8}$.

The motion-traced QFIs were evaluated using the same spectral-support
criterion stated above. The validation curves in
Fig.~\ref{fig:full_model_validation} use
$(n_f,n_m)=(42,9)$ and $181$ uniformly spaced times over
$0\le\tau\le12\pi$. The relative discrepancy $\varepsilon_{ac}$ is
evaluated only where its denominator is nonzero.

%%%%%%%%%%%%%%%%%%%%%%%%%%%%%%%%%%%%%%%%%%%%%%%%%%%%%%%%%%%%%%%%%%%%%
%%%%%%%%%%%%%%%%%%%%%%%%%%%%%%%%%%%%%%%%%%%%%%%%%%%%%%%%%%%%%%%%%%%%%
\section{Open-system Liouvillian and mixed-state QFI}
\label{app:liouvillian}
%%%%%%%%%%%%%%%%%%%%%%%%%%%%%%%%%%%%%%%%%%%%%%%%%%%%%%%%%%%%%%%%%%%%%

The Lindblad equation in Eq.\eqref{eq:lindblad} may be
vectorized by column stacking. The basic vectorization identities and the
Hamiltonian part of the Liouvillian are
\begin{equation}
{
\begin{aligned}
        \mathrm{vec}(A\rho B)
        &=
        (B^T\otimes A)\mathrm{vec}(\rho),
        \\
        \mathrm{vec}(H\rho)
        &=
        (\mathbb I\otimes H)\mathrm{vec}(\rho),
        \qquad
        \mathrm{vec}(\rho H)
        =
        (H^T\otimes\mathbb I)\mathrm{vec}(\rho),
        \\
        \mathrm{vec}([H,\rho])
        &=
        (\mathbb I\otimes H-H^T\otimes\mathbb I)\mathrm{vec}(\rho),
        \\
        \mathcal L_H
        &=
        -\frac{i}{\hbar}
        \left(\mathbb I\otimes H-H^T\otimes\mathbb I\right).
\end{aligned}
}
\end{equation}
For the dissipator,
\begin{equation}
{
\begin{aligned}
\mathrm{vec}(L\rho L^\dagger)
&=
(L^*\otimes L)\mathrm{vec}(\rho),
\\
\mathrm{vec}(L^\dagger L\rho)
&=
(\mathbb I\otimes L^\dagger L)\mathrm{vec}(\rho),
\\
\mathrm{vec}(\rho L^\dagger L)
&=
((L^\dagger L)^T\otimes\mathbb I)\mathrm{vec}(\rho).
\end{aligned}
}
\end{equation}
Thus
\begin{equation}
{
\begin{aligned}
\mathcal L
&=
-\frac{i}{\hbar}
\left(\mathbb I\otimes H-H^T\otimes\mathbb I\right)
\\
&\quad+
\sum_j\left[
L_j^*\otimes L_j
-\frac12\left(\mathbb I\otimes L_j^\dagger L_j
+(L_j^\dagger L_j)^T\otimes\mathbb I\right)
\right].
\end{aligned}
}
\end{equation}
For physical time and dimensionless time, respectively, the jump
operators are
\begin{equation}
{
\begin{aligned}
        L_1&=\sqrt{\kappa_{\rm ph}}\,a,
        &
        \widetilde L_1&=\sqrt{\tilde\kappa}\,a,
        \\
        L_2&=\sqrt{\gamma_{\rm ph}}\,\sigma_-,
        &
        \widetilde L_2&=\sqrt{\tilde\gamma}\,\sigma_-,
        \\
        L_3&=\sqrt{\gamma_{\phi,{\rm ph}}/2}\,\sigma_z,
        &
        \widetilde L_3&=\sqrt{\tilde\gamma_\phi/2}\,\sigma_z.
\end{aligned}
}
\end{equation}
The dimensionless Hamiltonian used in the $\tau=\Gmax t$
evolution is
$\widetilde H={H}/{\hbar\Gmax}.$ The vectorized physical-time and dimensionless-time solutions
are
\begin{equation}
{
        \mathrm{vec}[\rho(t;\phi)]
        =
        e^{\mathcal Lt}\mathrm{vec}[\rho(0)],
        \qquad
        \mathrm{vec}[\rho(\tau;\phi)]
        =
        e^{\widetilde{\mathcal L}\tau}\mathrm{vec}[\rho(0)].
}
\end{equation}
For the numerical QFI, the density matrix and its parameter tangent are propagated together. With $H(\phi)/(\hbar\Gmax)=\cos(\theta_0-\phi)K$ and $\partial_\phi H/(\hbar\Gmax)=\sin(\theta_0-\phi)K$, the augmented evolution is
\begin{equation}
\frac{d}{d\tau}
\begin{pmatrix}
\mathrm{vec}\rho\\[2pt]
\mathrm{vec}(\partial_\phi\rho)
\end{pmatrix}
=
\begin{pmatrix}
\widetilde{\mathcal L} & 0\\
\partial_\phi\widetilde{\mathcal L} & \widetilde{\mathcal L}
\end{pmatrix}
\begin{pmatrix}
\mathrm{vec}\rho\\[2pt]
\mathrm{vec}(\partial_\phi\rho)
\end{pmatrix}.
\label{eq:tangent_liouvillian}
\end{equation}
This construction preserves $\Tr(\partial_\phi\rho)=0$ up to numerical roundoff and avoids subtracting independently propagated density matrices.
The mixed-state QFI is then evaluated from
Eq.\eqref{eq:mixed_QFI}. The loss rates are reported as
tilded dimensionless ratios. The guide lines in Fig.~\ref{fig:open}(b) are
least-squares fits over $\tilde\kappa\in[0.05,0.5]$ and are used only to
summarize the displayed finite-range data.

\end{document}